\documentclass[a4paper,fleqn,10pt]{article}
\pdfoutput=1
\usepackage{amsmath}
\usepackage{amssymb}
\usepackage{array}
\usepackage{calc}
\usepackage{longtable}
\usepackage{multirow}
\usepackage[T1]{fontenc}
\usepackage{pstricks}
\usepackage{graphicx}
\usepackage{xspace}
\usepackage{units}

\numberwithin{equation}{section}
\usepackage{mciteplus}
\usepackage[pdfborder={0 0 0}]{hyperref}
\usepackage[format=hang,labelfont=bf,hypcap=true]{caption}
\usepackage{subcaption}
\usepackage{sectsty}
\usepackage{enumitem}
\allsectionsfont{\sffamily}
\subsubsectionfont{\mdseries\itshape\large}
\makeatletter
\DeclareRobustCommand*{\bfseries}{%
  \not@math@alphabet\bfseries\mathbf
  \fontseries\bfdefault\selectfont
  \boldmath
}
\makeatother
\let\spreprint\empty
\newcommand{\preprint}[1]{\def\spreprint{\protect#1}}
\let\sinstitute\empty
\newcommand{\institute}[1]{\def\sinstitute{\protect#1}}
\makeatletter
\renewcommand{\maketitle}{\begingroup
  \null\thispagestyle{empty}%
    \ifx\spreprint\empty
      \vskip 5ex
    \else
      \flushright\large\spreprint\vskip 10ex
    \fi
    \vskip 5ex
    \flushleft
      {\sffamily\bfseries\huge\@title}\vskip 6ex
      \@author\vskip 2ex
      \ifx\sinstitute\empty
      \else
        {\small\sinstitute}
      \fi
    \vskip 5ex
  \endgroup
}
\makeatother
\renewenvironment{abstract}{\begin{center}
  {\large\sffamily\bfseries Abstract: }
  \begin{minipage}[t]{0.75\textwidth}
}{\end{minipage}\end{center}\vskip 10ex}

\numberwithin{equation}{section}
\allowdisplaybreaks[2]

\newcommand{\ie}{\textit{i.e.\@}\xspace}

\newcommand{\see}{\textit{see\@}\xspace}

\newcommand{\Powheg}{P\protect\scalebox{0.8}{OWHEG}\xspace}

\newcommand{\Herwig}{H\protect\scalebox{0.8}{ERWIG}\xspace}

\newcommand{\Pythia}{P\protect\scalebox{0.8}{YTHIA}\xspace}

\newcommand{\Photos}{P\protect\scalebox{0.8}{HOTOS}\xspace}

\newcommand{\Horace}{H\protect\scalebox{0.8}{ORACE}\xspace}
\newcommand{\Winhac}{W\protect\scalebox{0.8}{INHAC}\xspace}
\newcommand{\Zinhac}{Z\protect\scalebox{0.8}{INHAC}\xspace}
\newcommand{\Rady}{R\protect\scalebox{0.8}{ADY}\xspace}

\newcommand{\Rivet}{R\protect\scalebox{0.8}{IVET}\xspace}

\newcommand{\Sherpa}{S\protect\scalebox{0.8}{HERPA}\xspace}

\newcommand{\Tevatron}{Tevatron\xspace}

\long\def\symbolfootnote[#1]#2{\begingroup%
\def\thefootnote{\fnsymbol{footnote}}\footnote[#1]{#2}\endgroup}

\newcommand{\im}{\imath}
\newcommand{\jm}{\jmath}
\newcommand{\ijt}{{\widetilde{\im\hspace*{-1pt}\jm}}}

\newcommand{\kt}{{\tilde{k}}}

\newcommand{\at}{{\tilde{a}}}

\newcommand{\ffbar}{\mathrm{f}\bar{\mathrm{f}}}

\newcommand{\done}{{\rm d}}
\newcommand{\order}{\mathcal{O}}

\newcommand{\bea}{\begin{eqnarray}}
\newcommand{\eea}{\end{eqnarray}}
\newcommand{\bi}{\begin{itemize}}
\newcommand{\ei}{\end{itemize}}

\newcommand*{\TeV}{\ensuremath{\text{Te\kern -0.1em V}}}
\newcommand*{\GeV}{\ensuremath{\text{Ge\kern -0.1em V}}}
\newcommand*{\MeV}{\ensuremath{\text{Me\kern -0.1em V}}}
\newcommand*{\keV}{\ensuremath{\text{ke\kern -0.1em V}}}
\newcommand*{\eV}{\ensuremath{\text{e\kern -0.1em V}}}

\newcommand{\tstart}{\ensuremath{t_\text{start}}\xspace}
\newcommand{\qbar}{\ensuremath{\bar{q}}\xspace}
\newcommand{\kT}{\ensuremath{{\mathrm{k}_\mathrm{T}}}\xspace}
\newcommand{\kTsq}{\ensuremath{{\mathrm{k}_\mathrm{T}^2}}\xspace}

\newcommand{\dRdress}{\ensuremath{\Delta \Theta_\text{dress}}}
\newcommand{\fdress}{\ensuremath{\mathrm{f}_\text{dress}}}
\newcommand{\ellp}{\ensuremath{\ell_\mathrm{p}}}
\newcommand{\ells}{\ensuremath{\mathrm{f}_\mathrm{s}}}

\newlist{myitemize}{itemize}{3}
\setlist[myitemize]{leftmargin=14em}

\newcolumntype{C}{>{\centering\arraybackslash}p{0.14\textwidth}}

\newlength{\unitcharwidth}
\hypersetup{
  pdfauthor={Lois Flower, Marek Schoenherr},
  pdftitle={Photon splitting corrections to soft-photon resummation}
}
\preprint{IPPP/22/69\\MCnet-22-19}
\author{Lois Flower, Marek Sch\"onherr}
\title{Photon splitting corrections to soft-photon resummation}
\institute{
  Institute for Particle Physics Phenomenology, Department of Physics, Durham University, Durham, DH1 3LE, UK
}
\begin{document}
\vspace*{10mm}
\maketitle
\vspace*{20mm}
\begin{abstract}
  In this paper we present an algorithm to add photon-splitting corrections
  to the Yennie-Frautschi-Suura-style soft-photon resummation available in
  the \Sherpa Monte-Carlo event generator.
  Photon-splitting corrections enter at NNLO in QED and, 
  as these effects are not incorporated in the
  standard QED FSR resummations, their size is larger than the pure hard
  photon-emission corrections at the same order.
  We introduce different lepton dressing strategies which incorporate 
  further leptons and hadrons in addition to the customary photons,
  and discuss their sensitivity to dressing parameters such as the 
  cone size.
  Finally, we quantify the effects of photon splittings into charged 
  fermions and scalars under
  different such dressing strategies on $Z\to e^+e^-$ decays and find
  effects of up to 1\% for suitably inclusive dressing strategies independent
  of the dressing cone size, and up to 9\% if only photons are used in the
  dressing procedure with large dressing cones.
\end{abstract}
\newpage
\tableofcontents
\section{Introduction}
\label{sec:intro}

Precision measurements of the Standard Model continue to stress-test
our understanding of particle physics at an unprecedented level.
In particular, charged and neutral Drell-Yan production at hadron colliders
like the LHC are used as standard candles due to their large cross
sections and exceedingly small experimental uncertainties, often
below the percent level.
However, these electroweak precision observables have also been 
brought to the forefront of searches for new physics, in the form of 
measured deviations from the Standard Model prediction. 
For example, the recent extraction of the $W$ boson mass, performed by the CDF 
experiment on legacy \Tevatron data \cite{CDF:2022hxs}, is in
apparent tension with the world average \cite{ParticleDataGroup:2022pth}
and previous hadron and lepton collider measurements \cite{LHCb:2021bjt,
  ATLAS:2017rzl,CDF:2013dpa,CDF:2013bqv,D0:2012kms,CDF:2012gpf,D0:2009yxq,
  CDF:2007mxw,CDF:2007cmy,CDF:2003tdi,CDF:2000gwd,D0:2002fhu,D0:1999rsi,
  D0:1999ivw,DELPHI:2008avl,ALEPH:2006cdc,L3:2005fft,OPAL:2005rdt},
as well as measurements of other fundamental EW parameters in $Z$ production
\cite{ATLAS:2016rnf,ATLAS:2012ewf,CMS:2015cyj,LHCb:2022tbc,CDF:2011ksg,
  CDF:2013uau,ALEPH:2005ab,CDF:2005qwt,ATLAS:2012au,CMS:2011kaj}.
Measurements such as this motivate 
precise theoretical input with uncertainties in the permille
range or lower.
At this level of required accuracy,  
higher-order QCD and electroweak corrections in vector-boson production 
must be supplemented with additional sources of theoretical precision.
In addition to a consideration of the structure functions describing 
the make-up of the incident particles, a detailed description of the
vector boson's decay is paramount.
Special emphasis must lie on the precise phase space distribution and flavour
composition of the accompanying radiation, in order to be able to precisely
model the detector response.
With this paper we contribute to the effort to determine the size and
uncertainty of higher-order QED corrections in the description of the
decay of massive vector bosons.

Higher-order corrections to Drell-Yan processes are known to
first order in the complete electroweak Standard Model
\cite{Wackeroth:1996hz,Baur:1997wa,Baur:1998kt,Baur:2001ze,Baur:2004ig,
  Andonov:2004hi,Dittmaier:2001ay,Dittmaier:2009cr,Li:2012wna,Alioli:2016fum}.
The recent advances at NNLO QCD-EW mixed calculations \cite{Bonciani:2021zzf,
  Buonocore:2021rxx,Armadillo:2022bgm,Delto:2019ewv,
  Buccioni:2020cfi,Bonciani:2020tvf,Bonciani:2021iis,Behring:2020cqi,
  Dittmaier:2014qza,Dittmaier:2015rxo},
though an impressive achievement in their own right, have not
increased the perturbative accuracy of the description of EW or
QED radiative corrections themselves.
Alternatively, universal QED corrections can be resummed to all orders
either in traditional QED parton showers \cite{Seymour:1991xa}
by means of the DGLAP equation,
or through the soft-photon resummation devised by Yennie, Frautschi,
and Suura (YFS) \cite{Yennie:1961ad}. These resummations can of course 
be matched to the fixed-order calculations mentioned above. 
A QED parton shower is available in all major Monte-Carlo event generators,
\Herwig \cite{Bellm:2019zci,Bellm:2015jjp},
\Pythia \cite{Bierlich:2022pfr,Sjostrand:2014zea}, and
\Sherpa \cite{Hoeche:2009xc,Gleisberg:2008ta,Bothmann:2019yzt}, while 
the YFS approach is implemented in \Herwig \cite{Hamilton:2006xz}
and \Sherpa \cite{Schonherr:2008av,Krauss:2018djz} for particle decays.
The implementation in \Sherpa has recently been extended to also
resum initial-state soft-photon radiation in $e^+e^-$ collisions
\cite{Krauss:2022ajk}.

To reach the necessary precision to make full use of the existing and
future experimental datasets, the QED effects impacting the leptonic
final state of the Drell-Yan process have to be understood in detail.
These effects are driven by soft and collinear photon radiation.
They can be resummed to all orders, and be further improved order by
order in perturbation theory.
Such calculations, matching to at least NLO EW corrections and sometimes
even including NNLO QED ones, have been implemented using QED parton showers
in \Horace \cite{CarloniCalame:2001ny,CarloniCalame:2003ux,
  CarloniCalame:2005vc,CarloniCalame:2006zq,CarloniCalame:2007cd,
  CarloniCalame:2016ouw}
and \Powheg \cite{Bernaciak:2012hj,Barze:2012tt,Muck:2016pko,Barze:2013fru},
using the structure function approach in \Rady
\cite{Dittmaier:2001ay,Dittmaier:2009cr}, and through a YFS-type
soft-photon resummation in \Winhac/\Zinhac \cite{Placzek:2003zg}, \Herwig
\cite{Hamilton:2006xz} and \Sherpa \cite{Schonherr:2008av,Krauss:2018djz}.
In addition, the \Photos Monte-Carlo \cite{Barberio:1990ms,
  Barberio:1993qi,Golonka:2005pn,Davidson:2010ew} provides an
algorithm based on both soft-photon resummation and matrix element corrections.
Dedicated comparisons between \Sherpa's YFS-type resummation and
\Photos \cite{Gutschow:2020cug}, between \Horace and \Photos
\cite{Kotwal:2015bfa}, as well as \Horace and \Winhac
\cite{CarloniCalame:2004fza} have yielded very good agreement.

A key element in the description of final state radiative corrections,
however, has only been sporadically and not very systematically addressed:
the possible splitting of the radiated bremsstrahlungs photons
into secondary charged-particle pairs.
These corrections only enter at a relative $\order(\alpha^2)$ in Drell-Yan
processes, but the production of light flavours may
be enhanced logarithmically and thus gain relevance.
In addition, and in contrast to QCD, photons and light charged flavours
like electrons, muons, or pions, are experimentally distinguishable --
such conversions alter the visible make-up of the final state and are thus
of importance at the envisaged theoretical precision.
It is also important to consider here the usual experimental and 
phenomenological practice of dressing charged leptons with photon radiation. 
While definitions of QCD jets have been constantly refined, there has 
been little discussion of dressed lepton algorithms since the adoption 
of cone-dressing strategies where all photons within a certain radius 
of the lepton are absorbed. Considering higher-order corrections in the 
form of photons splitting into charged particles has the potential to 
spoil the physically meaningful definition of a lepton dressed with 
photons. The treatment of charged leptons in the presence of secondary 
charged flavours must therefore be handled with care. 
Thus, while a first implementation of pair-production
corrections exists in \Photos \cite{Arbuzov:2012dx,Antropov:2017bed}, 
it only covers photon splittings into electrons and
muons, and their theoretical and phenomenological impact has not been
rigorously appraised.
This paper addresses this issue by introducing a rigorous independent
framework to calculate these corrections and study the resulting theoretical
and phenomenological implications.

This paper proceeds as follows: We begin by providing a brief summary
of the YFS soft-photon resummation as implemented in \Sherpa before
providing a comprehensive description of the photon splitting implementation,
including a detailed examination of their interplay and the splitting
properties in Sec.\ \ref{sec:methods}.
We then present a detailed discussion of possible extensions of the
standard lepton dressing algorithm to cope with the presence of
secondary pairs of (light) charged particles, and 
quantify their effect on $Z\to e^+e^-$ decays in Sec.\ \ref{sec:dress}.
Finally, we offer some concluding remarks in Sec.\ \ref{sec:conclusions}.

\section{Soft-photon resummation and photon splittings}
\label{sec:methods}

Incorporating photon-splitting processes alongside photon
emissions are straightforwardly implemented when both are described
in a common parton shower framework.
We prefer, however, to base our
implementation on the existing and superior description of
photon emission corrections in the YFS framework of \cite{Schonherr:2008av},
including its inherent coherent-radiation formulation and
existing NNLO QED and NLO EW corrections \cite{Krauss:2018djz}.
In this section we thus start by providing a brief summary of the
Yennie-Frautschi-Suura (YFS) soft-photon resummation and its
implementation in the \Sherpa event generator.
The remainder of this section then discusses the construction
of photon splitting algorithm in detail
before examining its properties.

\subsection{The YFS soft-photon resummation}
\label{sec:methods:yfs}

The work of Yennie\textendash{}Frautschi\textendash{}Suura (YFS) \cite{Yennie:1961ad} 
describes the infrared singularities of QED to all orders.
To achieve this, YFS consider all charged particles of the theory 
to be massive, and as a consequence only singularities associated with 
soft-photon emission are present.
In particular, all photon splittings are finite and thus do not
partake in the analysis of the infrared singular structure.
Using that knowledge, the YFS algorithm reorders the perturbative
expansion of a scattering or decay matrix element. This amounts 
to a resummation of the respective soft-photon logarithms in the 
enhanced real and virtual regions, leaving a perturbative expansion
in infrared-finite, hard photons (both real emisssions 
and virtual exchanges).

In the implementation of the YFS resummation in \Sherpa for
particle decays \cite{Schonherr:2008av}, the all-orders
soft-photon resummed differential decay rate is written as
\begin{equation}\label{eq:methods:yfsmaster}
  \begin{split}
    \done\Gamma^\text{YFS}
    =\;&
      \done\Gamma_0\cdot e^{\alpha Y(\omega_\text{cut})}\cdot
      \sum\limits_{n_\gamma=0}^\infty\frac{1}{n_\gamma!}
      \left[
        \prod\limits_{i=1}^{n_\gamma}\done \Phi_{k_i}\cdot\alpha\,
        \tilde{S}(k_i)\,\Theta(k_i^0-\omega_\text{cut})
        \cdot\mathcal{C}
      \right]\,,
  \end{split}
\end{equation}
wherein $\done\Gamma_0$ is the leading-order (LO) differential decay rate and
the YFS form factor $Y(\omega_\text{cut})$ contains the soft-photon
logarithms.
The decay rate is then summed over all possible additional photon
emissions with an energy larger than $\omega_\text{cut}$ wrt.\
the leading-order decay.
Each emission is described through its eikonal $\tilde{S}$ and
corrected for hard emission effects up to a given order through
the correction factor $\mathcal{C}$.\footnote{
  The hard (real and virtual) photon-emission corrections $\mathcal{C}$
  are available up to NLO EW for leptonic $W$ decays and up to
  NNLO QED + NLO EW for leptonic $Z$ decays \cite{Krauss:2018djz}.
}

Unlike a conventional parton shower, where the resummation
is reliant on the factorisation of subsequent emissions when ordered in an
evolution variable, YFS photons are unordered.
In addition, they are also emitted coherently from the charged multipole
through the radiator function $\tilde{S}$ and are thus not inherently
associated with a specific emitter particle.
Consequently, when the produced final state is to be further treated by
a dedicated photon-splitting parton shower, the existing configuration
must be interpreted in the parton shower's evolution and splitting
language before any further splittings take place.
Of course, care has to be taken so as to not compromise its leading logarithmic soft-photon resummation.
In the following algorithm, since the effects added 
are completely beyond the scope of the YFS formulation
without any potential overlap, this requirement amounts to ensuring the kinematic
recoil induced by a splitting photon on the primary charged particle ensemble
(and possibly other existing photons), vanishes in the limit that the
energy of the splitting photon vanishes.
While this is trivially true as all charged particles are treated
as massive, the recoil assignment performed in this study and described 
in section \ref{sec:methods:photonsplit} introduces corrections to the momenta of the primary 
charged particle ensemble which scale non-logarithmically with the photon 
energy and hence do not contribute to the leading-logarithmic resummation.

\subsection{Photon splittings}
\label{sec:methods:photonsplit}

In this section we introduce the parton shower algorithm
which computes the photon splitting probabilities and kinematics,
while the principal user input commands to steer its behaviour
are described in App.\ \ref{app:settings}.
We will use the usual notation associated with a Catani-Seymour
dipole shower, following \cite{Schumann:2007mg}.
Since the YFS algorithm requires massive charged particles,
it is necessary to include all masses in this shower for consistency.
There are therefore no infrared singularities associated with
our photon splittings, since the collinear pole is regulated
by the fermion and scalar masses.
However, the aim is still to capture the correct behaviour in
the quasi-collinear limit, accounting for the logarithmic
enhancement for collinear splitting into light flavours.
Throughout this section we will focus on configurations where
all relevant particles are in the final state of the
decay process, \ie decays of neutral resonances.
Decays of charged resonances are handled similarly;
the corresponding modifications are detailed in App.\ \ref{app:fidip}.

The key part of the parton shower algorithm is, as usual, the
veto algorithm \cite{Seymour:1994df,Sjostrand:2006za}.
This allows us to avoid the
problem of analytically integrating the splitting functions,
which are detailed below, by using an
overestimate to evaluate the cumulative emission probability,
and then vetoing emissions
with a probability which corrects for the overestimate.
The evolution begins at some starting scale $\tstart$ which is the 
highest possible scale for a splitting to take place;
we postpone its exact definition to the end of this section.
All splitting functions compete: a splitting scale is calculated
for each possible combination of splitter $\ijt$, splitting
products $i$ and $j$, and spectator $\kt$/$k$ (before/after
the splitting process).
Whichever splitting process would happen at the highest scale is selected.
If the splitting is accepted (not vetoed), a new particle is
created and flavours and kinematics of existing particles are updated.
The whole process is repeated, starting from the selected splitting scale, 
and iterated until some infrared cutoff $t_0$ is reached. 
This cutoff is needed to regulate the divergence
of the splitting functions in the general case
where these appear.
For a QCD shower, a physical choice for the cutoff is the
hadronisation scale, which is of $\order(1\,\text{GeV})$,
well above $\Lambda_\text{QCD}$ where QCD dynamics
turn non-perturbative.
For a QED shower, however, the splittings
which do not involve quarks can evolve to arbitrarily low scales.
In the algorithm presented here,
which contains only splitting functions of photon emissions off
charged scalars and fermions as well as of photons splitting into massive
fermions or pseudo-scalar hadrons,
the cutoff is dictated by the mass of the lightest
fermion, $t_0 = 4m_e^2$ or lower.

As stated earlier, in the case of a photon splitting to a fermion or 
scalar particle-antiparticle pair, there is no soft divergence.
The collinear divergence present for massless splitting products
is converted into a logarithmic collinear
enhancement when masses are included;
hence, lighter particles will have a larger contribution
to photon splitting corrections.
Here we include all possible splittings up to a mass cutoff of
$2 m_i \lesssim 1 \,\GeV$ in addition to $\tau$ pair production which,
while rare, contributes
to some observables through the decays to lighter leptons or hadrons.
Since most splittings occur
near or below the hadronisation scale, we consider hadrons, not quarks,
to be the relevant QCD degrees
of freedom.
Using this mass cutoff, the hadrons which can be produced are the charged pions
and kaons. They are pseudo-scalars, and their interaction with photons
is modeled using point-like scalar QED, neglecting any substructure
effects.
We use the scalar splitting functions of \cite{Catani:2002hc}.
Depending on the experimental environment, the kaons and $\tau$ leptons
decay before hitting any detector.
This can be handled within the usual (hadron) decay treatment available
within the \Sherpa framework \cite{Bothmann:2019yzt,Hoche:2014kca}.

\paragraph*{Splitting functions and spectator assignment.}
In the usual parton shower notation, we use the following
dipole splitting functions \cite{Catani:2002hc,Schumann:2007mg,Schonherr:2017qcj}
\begin{equation} \label{eq:methods:SFs}
  \begin{split}
    S_{s_\ijt(\kt)\to s_i\gamma_j(k)}
    \,=&\;
    S_{\bar{s}_\ijt(\kt)\to \bar{s}_i\gamma_j(k)}
    \,=\;
      -\,\mathbf{Q}^2_{\ijt\kt}\;\alpha\,
      \left[\frac{2}{1-z+zy}-\frac{\tilde{v}_{\ijt,\kt}}{v_{ij,k}}\left(2+\frac{m_i^2}{p_ip_j}\right) \right] \\
    S_{f_\ijt(\kt)\to f_i\gamma_j(k)}
    \,=&\;
    S_{\bar{f}_\ijt(\kt)\to \bar{f}_i\gamma_j(k)}
    \,=\;
      -\,\mathbf{Q}^2_{\ijt\kt}\;\alpha\,
      \left[\frac{2}{1-z+zy}-\frac{\tilde{v}_{\ijt,\kt}}{v_{ij,k}}\left(1+z+\frac{m_i^2}{p_ip_j}\right) \right] \\
    S_{\gamma_\ijt(\kt)\to s_i\bar{s}_j(k)}
    \,=&\;
    S_{\gamma_\ijt(\kt)\to f_i\bar{f}_j(k)}
    \,=\;
      -\,\mathbf{Q}^2_{\ijt\kt}\; \alpha\,
      \left[
        1-2z(1-z)-z_+ z_- \vphantom{\frac{m_i^2}{p_ip_j}}
      \right]
  \end{split}
\end{equation}
for splittings involving the scalars $s$, fermions $f$, their antiparticles
$\bar{s}$ and $\bar{f}$, and a photon $\gamma$,
in terms of the splitting variable $y$ and light-cone momentum fraction $z$.
These are defined as
\begin{equation} \label{eq:methods:zy}
  y = \frac{p_i p_j}{p_i p_j + p_i p_k + p_j p_k}
  \qquad\text{and}\qquad
  z = \frac{p_i p_k}{p_i p_k + p_j p_k}
  \;.
\end{equation}
Further, $m_i$ is the mass of the splitting product $i$, and
$z_-$ and $z_+$ are the phase space boundaries
\begin{equation} \label{eq:methods:zlimits}
  z_\pm = \frac{2 \mu_i^2 + (1-\mu_i^2-\mu_j^2-\mu_k^2)\,y}{2(\mu_i^2+\mu_j^2 + (1-\mu_i^2-\mu_j^2-\mu_k^2)\,y)} (1 \pm v_{ij,i} \, v_{ij,k})\;,
\end{equation}
where the dimensionless rescaled masses $\mu_i^2=m_i^2/Q^2$ are introduced for convenience, and
$Q^2=(p_i+p_j+p_k)^2=(p_\ijt+p_\kt)^2$ is the invariant mass of the dipole.
The relative velocities $\tilde{v}_{\ijt,\kt}$, $v_{ij,k}$, and $v_{ij,i}$
are given by
\begin{equation} \label{eq:methods:relv}
  \begin{split}
    \tilde{v}_{\ijt,\kt} \,=&\; \frac{\sqrt{\lambda(1,\mu_{\ijt}^2,\mu_\kt^2)}}{1-\mu_{\ijt}^2-\mu_\kt^2}\;, \\
    v_{ij,i} \,=&\; \frac{\sqrt{(1-\mu_i^2-\mu_j^2-\mu_k^2)^2 \,y^2 -4\mu_i^2\mu_j^2}}{(1-\mu_i^2-\mu_j^2-\mu_k^2)\,y+2\mu_i^2}\;, \\
    v_{ij,k} \,=&\; \frac{\sqrt{(2\mu_k^2+(1-\mu_i^2-\mu_j^2-\mu_k^2)(1-y))^2-4\mu_k^2}}{(1-\mu_i^2-\mu_j^2-\mu_k^2)(1-y)}\;.
  \end{split}
\end{equation}
Finally, the charge correlator $\mathbf{Q}^2_{\ijt\kt}$ is defined as
\cite{Yennie:1961ad,Dittmaier:2008md,Kallweit:2017khh,Schonherr:2017qcj}
\begin{equation} \label{eq:methods:chargecorrelator}
  \mathbf{Q}^2_{\ijt\kt} = \begin{cases}
     \;\;\frac{Q_{\ijt} Q_{\kt} \theta_{\ijt} \theta_{\kt}} {Q^2_{\ijt}} & \ijt \neq \gamma \\
     \;\;\kappa_{\ijt\kt} & \ijt=\gamma
  \end{cases}
  \qquad\text{with}\qquad
  \sum\limits_{\kt\neq\ijt}\kappa_{\ijt\kt}=-1\quad\forall\,\ijt=\gamma\;,
\end{equation}
where the $Q_{\ijt}$ and $Q_{\kt}$ are the charges of the splitter and
spectator respectively and their $\theta_{\ijt/\kt}$ are 1 ($-1$) if
they are in the final (initial) state.
The $\kappa_{\gamma\kt}$ need to ensure that the splitting
functions are appropriately normalised such that the correct collinear
limit is found, but are otherwise unconstrained.
Here we choose
\begin{equation}
  \kappa_{\gamma\kt} = -\frac{1}{\mathcal{N}_\mathrm{specs}},
\end{equation}
where $\mathcal{N}_\mathrm{specs}$ is the chosen number of possible
spectators, \ie we choose to weigh all selected spectators $\kt$ equally.
The photon splittings themselves are free of soft divergences, hence the
spectator is only needed for momentum conservation and, in
principle, any other particle of the process may assume this role.
In the present context, we consider all primary charged decay products
as possible spectators of photon splittings as our default, but the choice
to consider only the splitting photon's originator particle (as
reconstructed, described below) has also been implemented, \see
App.\ \ref{app:settings}.
While the other present YFS photons and other neutral decay products
as well as the decaying particle itself are all valid spectators, the two
choices described above both guarantee that enough energy is available 
to allow photon splitting into heavier flavours to occur.
Limiting the number of spectators also helps to reduce the computational
complexity.
For photon radiation off the products of a photon splitting,
the spectator assignment, and therefore the recoil, is kept
local in that system.

\paragraph*{Evolution variable.} \label{par:methods:evovariable}
For the evolution variable $t$ used in the parton 
shower, the requirement of leading logarithmic accuracy 
means that any choice which preserves $\mathrm{d}t/t$ is formally  
equivalent in the infrared limit.
In a given splitting function we consider two variants,
virtuality $\qbar^2$ and
transverse momentum $\kTsq$.
The virtuality is defined, in terms of the dipole invariant mass $Q^2$
and the masses of the emitter $m_\ijt$, splitting products $m_{i/j}$
and spectator $m_k$, as 
\begin{equation}\label{eq:methods:startingscalevirt}
  \qbar^2 = (Q^2-m_i^2-m_j^2-m_k^2)\,y + m_i^2 + m_j^2 - m_{\ijt}^2\;.
\end{equation}
Specifically, for the two relevant cases this translates to
\begin{equation}\label{eq:methods:startingscalevirt1}
  \qbar_{\mathrm{f}\to\mathrm{f}\gamma}^2
  = (Q^2-m_\mathrm{f}^2-m_k^2)\,y
\end{equation}
for photon emissions, where the flavour $\mathrm{f}$
can either be a scalar $s$, fermion $f$, or their
antiparticles $\bar{s}$ and $\bar{f}$, and
\begin{equation}\label{eq:methods:startingscalevirt2}
  \qbar_{\gamma\to\mathrm{f}\bar{\mathrm{f}}}^2
  = (Q^2-2m_\mathrm{f}^2-m_k^2)\,y + 2m_\mathrm{f}^2
\end{equation}
for photon splittings.
We see that, as stated earlier, photon emissions are possible at arbitrarily
low evolution scales while photon splittings can only occur
if the virtuality exceeds the pair-production threshold.
Likewise, the transverse momentum is defined by \cite{Schumann:2007mg}
\begin{equation}\label{eq:methods:startingscaleKt}
  \kTsq = (Q^2-m_i^2-m_j^2-m_k^2)\,y\,z(1-z) - m_i^2\,(1-z)^2 - m_j^2\,z^2\;.
\end{equation}
Again, for photon emissions this translates to
\begin{equation}\label{eq:methods:startingscaleKt1}
  \kTsq_{\,\mathrm{f}\to\mathrm{f}\gamma}
  = (Q^2 - m_\mathrm{f}^2-m_k^2)\,y\,z(1-z) - m_\mathrm{f}^2\,(1-z)^2
\end{equation}
and to
\begin{equation}\label{eq:methods:startingscaleKt2}
  \kTsq_{\,\gamma\to\mathrm{f}\bar{\mathrm{f}}}
  = (Q^2 - 2m_\mathrm{f}^2-m_k^2)\,y\,z(1-z) - m_\mathrm{f}^2\,(z^2+(1-z)^2)\;
\end{equation}
for photon splittings.
As discussed, photon emissions are possible down to $\kTsq=0$,
but in this case the photon-splitting threshold also lies at $\kTsq=0$.
As a result, the chosen infrared cutoff $t_0$ will induce a minimal
$\kT$, and thus opening angle, produced in the pair-creation process.
Hence, in full analogy to most QCD parton showers, the pair's
virtuality $\qbar^2$, with its automatic introduction of a pair-production
threshold, is expected to give a better description.


It can be seen that the relation 
\begin{equation}\label{eq:methods:psdifferential}
  \frac{\mathrm{d}t}{t} = \frac{\mathrm{d}\kTsq}{\kTsq} = \frac{\mathrm{d}\bar{q}^2}{\bar{q}^2}
\end{equation}
holds, therefore both transverse momentum and virtuality are possible choices
of evolution variable and no Jacobian is needed to translate between them.

Using these definitions, there are three well-motivated choices for
the global evolution variable.
\begin{enumerate}
 \item
    As in most QCD parton showers, $t=\kTsq$ is a viable choice.In analogy to QCD,
    ordering photon emissions by transverse momentum
    results in the inclusion of charge coherence effects \cite{Amati:1980ch},
    but there is no particular motivation to use $t=\kTsq$ as the evolution
    variable for photon splittings into charged-particle pairs.
  \item
    Choosing $t=\qbar^2$ is an equally valid option.
    Due to the $s$-channel nature of photon splittings, the photon virtuality
    is expected to be a good ordering variable here \cite{Bierlich:2022pfr,
    Brodsky:1982gc}.
    But since it does not implement angular ordering natively, it is not
    expected to yield the best description of soft-photon emissions.
  \item
    Following eq.\ \eqref{eq:methods:psdifferential},
    we are free to interpret the evolution variable differently in
    different splitting processes as long as $\done t/t$ is invariant.
    As our default, we thus choose to interpret the evolution
    variable $t$ as $\kTsq$ in photon emissions and as $\qbar^2$
    in photon splittings. We will call this the ``mixed scheme'' in
    later sections.
\end{enumerate}
All three choices are implemented, \see App.\ \ref{app:settings},
and some of their respective consequences will be explored in
Sec.\ \ref{sec:methods:diagnostics}.

\paragraph*{Generation of splitting variables.}
While the evolution variable $t$ is generated as usual in the veto
algorithm, the light-cone momentum fraction $z$ has to be generated within
its allowed range $[z_-,z_+]$.
The integration limits $z_\pm$ are defined in 
eq.\ \eqref{eq:methods:zlimits}, but in order
to generate a Sudakov factor we work with the evolution variable $t$.
We generate a trial emission using the integral of the
overestimate of the splitting function, for which the
$z$ limits are necessary, but
we do not yet know the value of the kinematic variable
$y$ (eq.\ \eqref{eq:methods:zy}).
Using a change of variables to replace 
$y$ with the evolution parameter yields usable $z$ ranges at this stage. 
Hence in the $\kT$ ordered scheme, the $z$ limits are \cite{Schumann:2007mg}
\begin{equation}
  z_{\pm,\kT} = \text{min/max}{\left[\frac{1}{2} \left( 1 \pm \sqrt{1-\frac{4t_0}{Q^2}} \right),\,z_\pm\right]}.
\end{equation}
Note that the $z_\pm$ are not yet known, but can be overestimated by 0 and 1,
respectively.
The above expression thus gives an overestimate of the true phase space
available.
The number of splittings rejected as a result is very small, however,
and does not have a large impact on the efficiency of the algorithm.

On the other hand, in the virtuality ordered scheme $\qbar^2$ has no $z$ dependence.
This means that $y$ can be determined independently of the
light-cone momentum fraction $z$ as well, $y(t,z)=y(t)$, by
solving eq.\ \eqref{eq:methods:startingscalevirt} for $y$.
This implies that the $z$ limits can be written as 
\begin{equation}
  z_{\pm,\qbar}
  = \text{min/max}
    {\left[
      \frac{q^2+m_i^2-m_j^2}{2q^2}
      \left(
        1
        \pm \sqrt{1-\frac{4m_i^2 q^2}{\left(q^2+m_i^2-m_j^2\right)^2}} \,
            \sqrt{1+\frac{4m_k^2q^2}{\left(Q^2-q^2-m_k^2\vphantom{m_j^2}\right)^2}}
      \right),\,z_\pm\right]}
\end{equation}
where $q^2=\qbar^2+m_\ijt^2$.
Again, however, it is only a small price in efficiency to use larger and simpler limits at 
the trial emission stage. In the results that follow, $z_{-,\bar{q}^2}=0$ and 
$z_{+,\bar{q}^2}=1$ have been used. 

\paragraph*{Kinematics.}
With the above definitions of the dipole variables $y$ and $z$ or, 
alternatively, with the evolution variable $t$ and the splitting variable $z$,
and the
uniformly distributed azimuthal splitting angle $\phi$, we can now
build the kinematics of the splitting products $i$ and $j$ and the
spectator $k$ after the emission process.
The new momenta are given by an inversion of the momentum maps
of \cite{Catani:2002hc}, and in their construction we follow
\cite{Schumann:2007mg}.
In particular, for the final-final (FF) dipoles discussed here,
they are given in Sec.\ 3.1.1 eq.\ (49)--(58) of \cite{Schumann:2007mg}.
Note that this redistribution of momenta is infrared safe and
does not spoil the leading logarithmic accuracy of the YFS resummation, since
it introduces non-logarithmic corrections only.

\paragraph*{Starting conditions.}
Having defined the evolution and splitting variables as well as
the splitting functions and kinematic mappings above, we now
need to specify the initial conditions to fully define the algorithm.
As the photon emissions are already generated by the YFS soft-photon
resummation, the existing distribution has to be reinterpreted as if
it was generated by our shower algorithm.
Then, the missing photon-splitting corrections can be embedded into
the existing calculation.
By not allowing further photon-radiation off the primary
charged-particle ensemble, double counting is avoided.

To determine the scale at which each existing photon has been
produced, we calculate the emission probabilities according to
the splitting functions $S_{f_\ijt(\kt)\to f_i\gamma_j(k)}$ and
$S_{s_\ijt(\kt)\to s_i\gamma_j(k)}$, respectively, for every
existing soft-photon $\gamma_j$.
Therein, every primary charged particle (all existing charged particles
of the process at the this stage) can act as possible emitter $\ijt$
and spectator $\kt$.
One of those possible splitting functions is then selected either
according to its probability $S_{\ijt\kt\to ijk}/\sum_{\ijt\kt} S$
(default), or by selecting the one with the largest splitting
probability (\see App.\ \ref{app:settings}).
Its reconstructed evolution variable $t$ is then set as the starting
scale $t_{\text{start},j}$ of the further evolution for photon $\gamma_j$.
The above parton shower algorithm is then started at the largest
of all photons' starting scale, $\tstart=\max[t_{\text{start},j}]$,
but each individual photon's evolution is only active for
$t\leq t_{j,\text{start}}$.

\subsection{Properties of the photon-splitting algorithm}
\label{sec:methods:diagnostics}

Having the algorithm to calculate photon splitting probabilities
at hand, we can now examine its properties and assess the
consequences of specific algorithmic choices discussed above.
To be precise, we use the example of an on-shell $Z$ boson
decaying to an $e^+e^-$ pair (maximising the number of radiated
photons).
Hence, as we are not in a collider environment, we use
a spherical coordinate system to measure relative radial
distances $\Delta\Theta$ in the following.

We begin by presenting a detailed look into the conditions under
which the photons generated through the YFS soft-photon resummation
split.
As discussed, in a first step, the existing distribution of photons
and primary emitters has to be clustered to assign individual
starting scales to the evolution of each photon.
This assignment is of course dependent on the choice of evolution
variable for photon emissions off charged particles as well as the
choice of spectator scheme.

\begin{figure}[t!]
  \centering
  \includegraphics[width=0.47\textwidth]{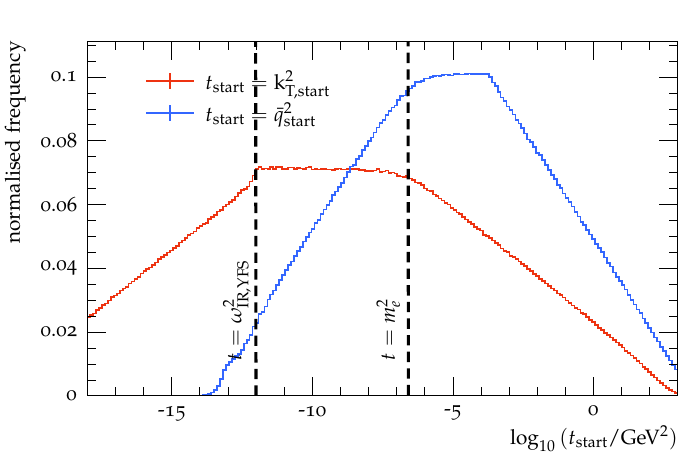}
  \hfill
  \includegraphics[width=0.47\textwidth]{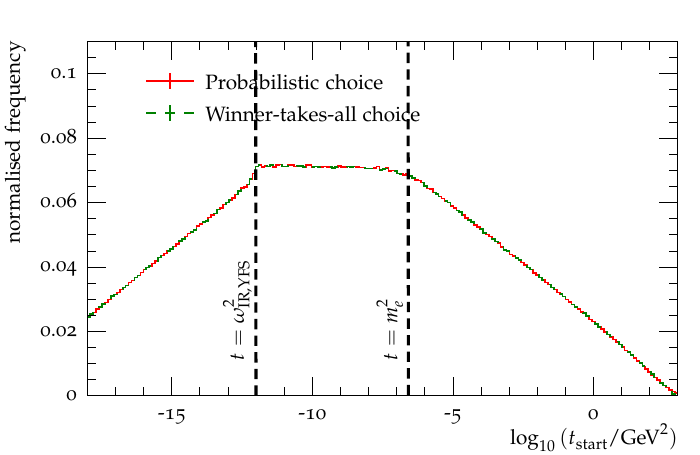}
  \caption{
    \textbf{Left:} A comparison of the frequency of the reconstructed starting scales \tstart
    using two choices for the evolution variable $t$, $\kTsq$ or $\qbar^2$,
    in the reconstructed initial $e^\pm\to e^\pm\gamma$ splitting.
    \textbf{Right:} A comparison of the frequency of the reconstructed starting scales
    $t_\text{start}=\mathrm{k}_{\mathrm{T},\text{start}}^2$
    using either a probabilistic determination of the emitter lepton
    or a winner-takes-all
    in the reconstructed initial $e^\pm\to e^\pm\gamma$ splitting.
    The threshold for photons splitting into charged particle pairs is
    $t>4m_e^2$, and $\omega_{\text{IR},\text{YFS}}^2$ is the infrared
    cutoff of the YFS-style algorithm which generates the photons.
    \label{fig:results:startingscale}
  }
\end{figure}

Fig.\ \ref{fig:results:startingscale} shows the distribution
of starting scales when the photon emissions are reconstructed with the inverse
emission kernels.
In the left plot, the ordering variable is interpreted as
either a relative transverse momentum, $t=\kTsq$ (red) or a
virtuality, $t=\qbar^2$ (blue).
In the transverse momentum ordering scheme we observe an approximately
logarithmic rise in the abundance of starting scales, starting at
the kinematic limit of $\kTsq\simeq\tfrac{1}{4}\,m_Z^2$.
This reflects the photon spectrum produced by the soft-photon resummation.
This logarithmic rise levels out at $\kTsq\approx m_e^2$, formed by
the reconstructed $\kTsq$ of ultra-soft photons of the event.
This plateau ends at the soft-photon cutoff $\omega^2_\text{IR,YFS}$
 used in the soft-photon resummation.
In contrast, in the virtuality ordering
scheme we see the majority of events have starting scales above $10^{-6}$ GeV.
In both cases the characteristic scale at $t=m_e^2$ is induced by both the splitter and the spectator masses of the primary decay.
The effect of the infrared cutoff is
straightforward in the $t=\kTsq$ case, as shown by the labelled black dashed
line. In the $t=\qbar^2$ case the cutoff does not dictate the turning point
of the frequency plot, but, indirectly, the point at which the frequency falls
to zero. Due to normalisation this has the effect of increasing the frequency
above the electron mass. This appears as a flattening off of the plot just above
the electron mass squared, before the frequency falls towards zero at $m_Z^2$
independent of the IR cutoff.
We note that in the mixed ordering scheme, which we choose as our default
ordering variable scheme, $\tstart = \mathrm{k}_{\mathrm{T},\text{start}}^2$.

On its right-hand side, Fig.\ \ref{fig:results:startingscale} shows
the distribution of starting
scales $\tstart=\kTsq$ when the reconstructed emission of a YFS photon from
one of the final-state charged particles is chosen either probabilistically
according to the relative sizes of the splitting functions (red, our default) or by simply
choosing the more likely emitter (green, dashed). There is no significant difference
between the two schemes. A very small difference occurs at the high $\tstart$ end.
In the winner-takes-all
scheme, large starting scales are less likely because the emitter with
the largest splitting function is always chosen; the chosen emitter is
the particle which results
in a smaller starting scale, due to the soft divergence and collinear enhancement
of the splitting functions.

\begin{figure}[t!]
  \centering
  \includegraphics[width=0.32\textwidth]{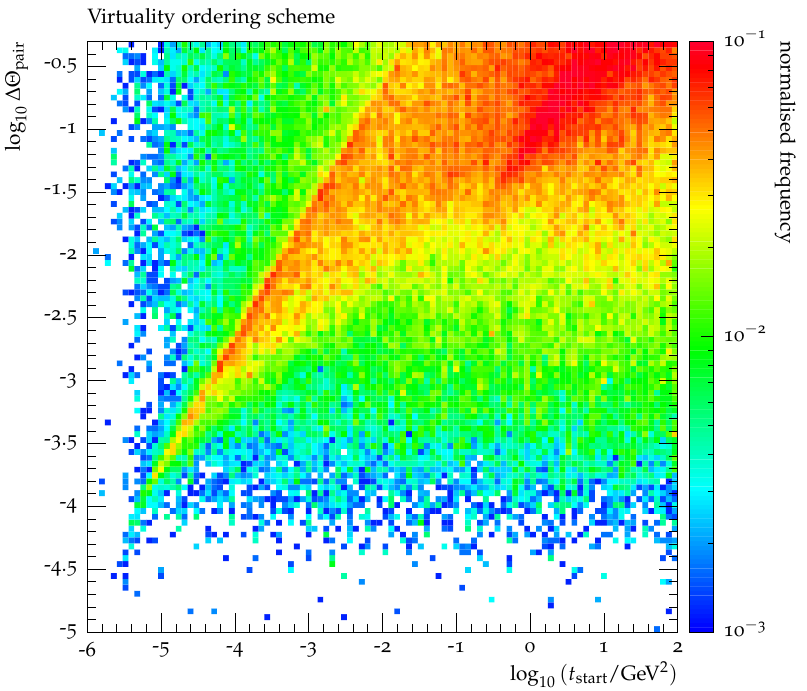}
  \includegraphics[width=0.32\textwidth]{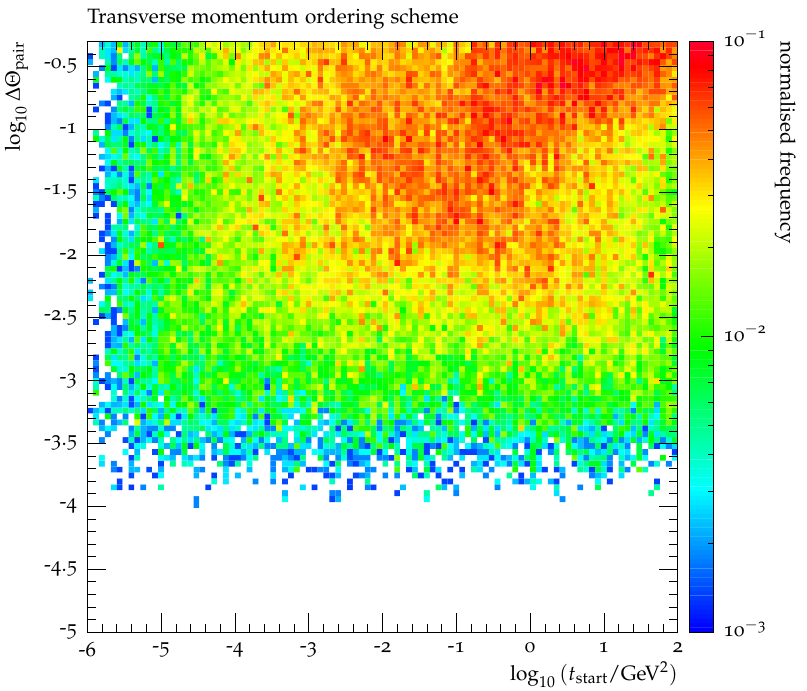}
  \includegraphics[width=0.32\textwidth]{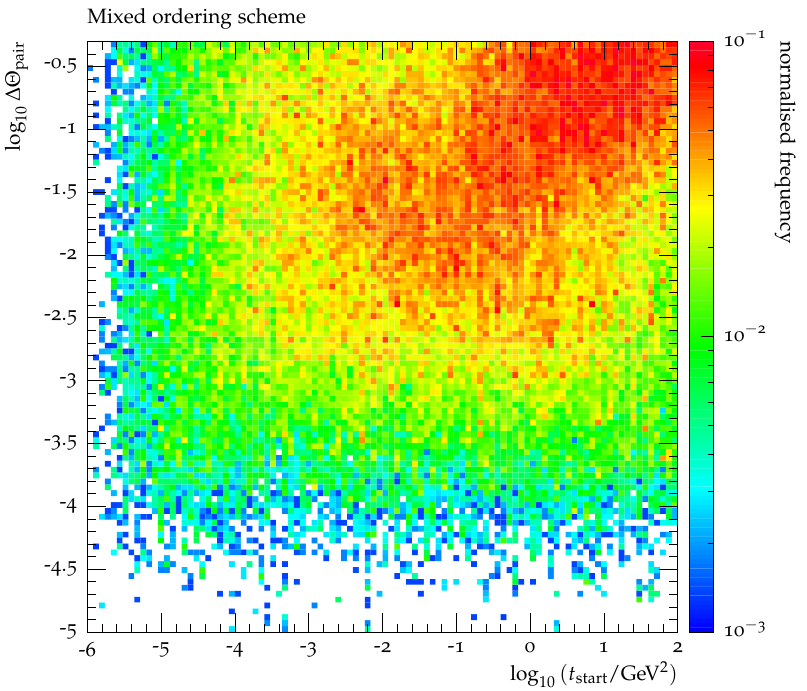}
  \caption{
    The interdependence of the starting scale $\tstart$ of a photon
    and the angular separation between the particles produced
    in its splitting, $\Delta \Theta_\mathrm{pair}$, in the
    in the $\qbar$-ordered scheme (left), the $\kT$-ordered
    scheme (centre), and the mixed ordering scheme (right).
    \label{fig:results:tdR}
  }
\end{figure}

Having established the starting conditions of the photons' evolution
we can now examine their splitting process into pairs of charged
particles, and the interplay of the choice of interpretation of the
evolution variable in either splitting process.
Therefore, Fig.\ \ref{fig:results:tdR} depicts the correlation of the
starting scale $t_{\text{start},j}$ of a photon and the collinearity,
or opening angle $\Delta\Theta_\text{pair}$, of 
its splitting products (mainly $e^+e^-$ pairs),
for all three different choices of interpretation of the evolution
variable $t$: $\qbar^2$, $\kTsq$, or mixed.
In the virtuality-ordered scheme, $t=\qbar^2$,
photons can only split if $t$ exceeds the pair-creation threshold of
the lightest charged species, $t\geq 4\,m_e^2\approx 10^{-6}\,\text{GeV}^2$.
Further, there is also a strong correlation between
the starting scale of the evolution and the eventual splitting angle,
which is mainly a consequence of the identification of the starting
scale $\tstart$ for each photon.
As already anticipated in section \ref{sec:methods:photonsplit} above,
the evolution scale in the $\kT$-ordered case is only constrained to be
above the infrared cutoff $t_0$, which is also chosen to be
$t_0=4\,m_e^2$ here.
This constrains the opening angle of the pair of splitting 
products to be $\Delta\Theta_\text{pair} \gtrsim 10^{-4}$.
The mixed scheme, interpreting $t=\kTsq$ in reconstructing the photon
emission to define $\tstart$ and $t=\qbar^2$ in photon splittings,
combines aspects of these two schemes, producing a smooth
distribution in the whole $(\tstart,\Delta \Theta_\text{pair})$ space independent
of $t_0$, as long as $t_0\leq 4\,m_e^2$.
The opening angle becomes relevant when studying the recombination
properties of the splitting product into a dressed primary charged
particle, \see Sec.\ \ref{sec:results:dress}.

\begin{figure}[t!]
  \centering
  \includegraphics[width=0.32\textwidth]{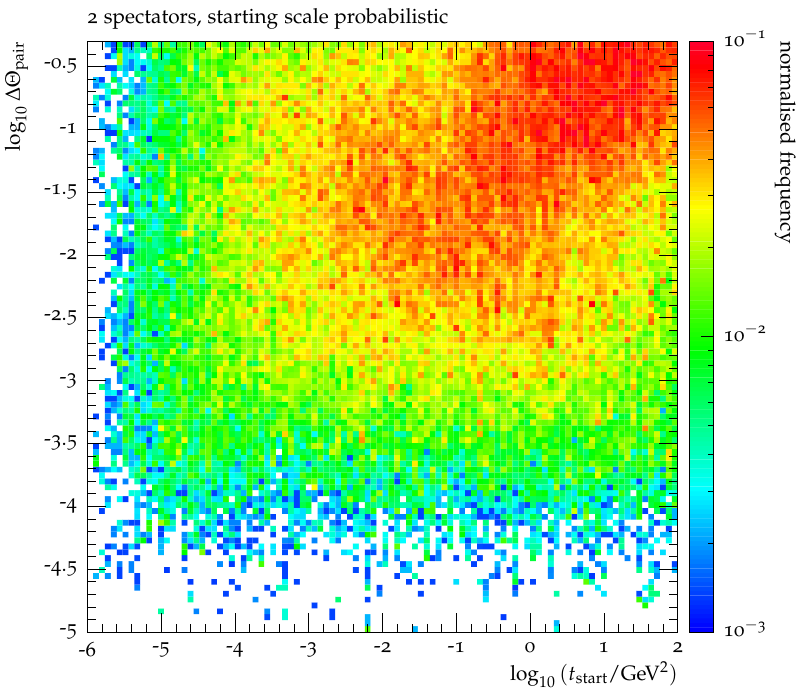}
  \hspace*{0.05\textwidth}
  \includegraphics[width=0.32\textwidth]{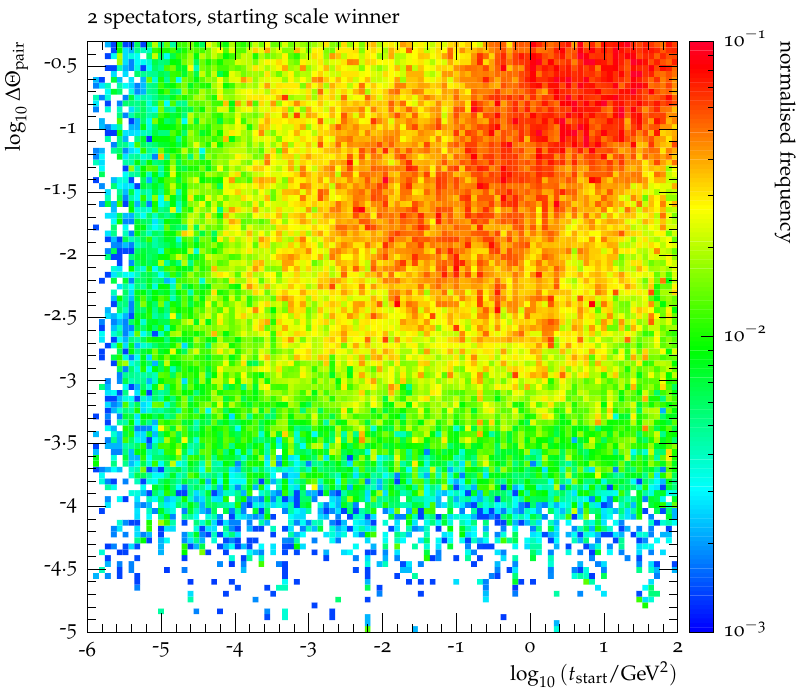}
  \\[2mm]
  \includegraphics[width=0.32\textwidth]{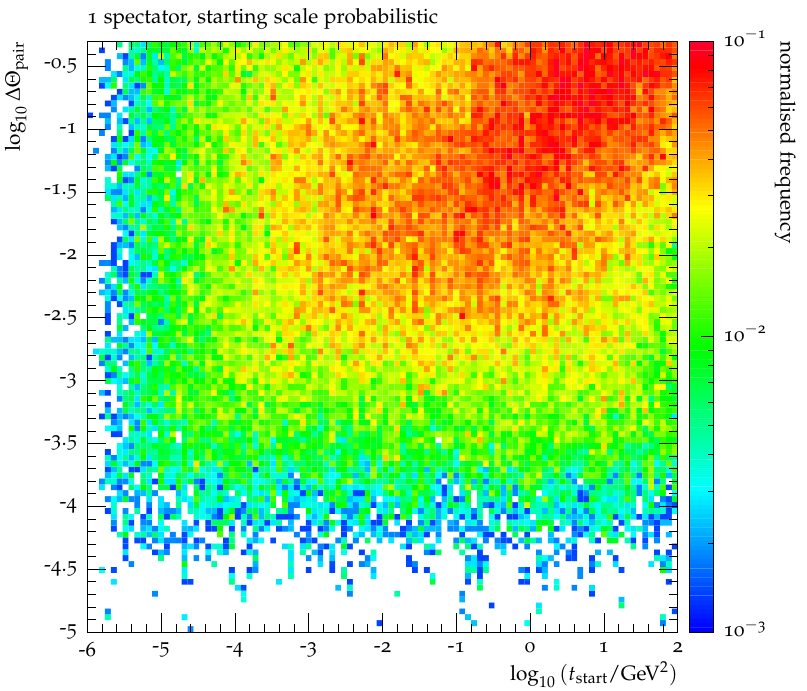}
  \hspace*{0.05\textwidth}
  \includegraphics[width=0.32\textwidth]{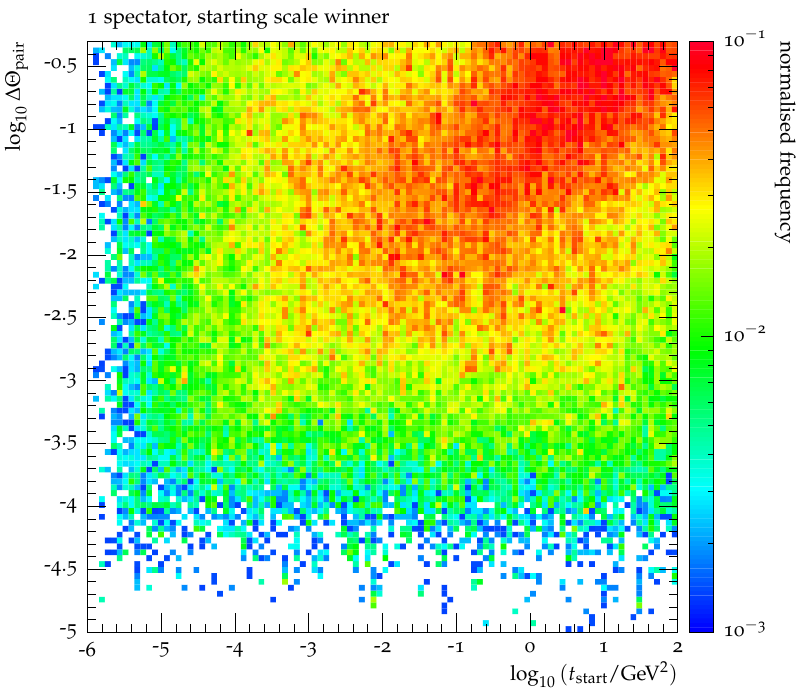}
  \caption{
    The interdependence of the starting scale $\tstart$ of a photon
    and the angular separation between the particles produced
    in its splitting, $\Delta \Theta_\mathrm{pair}$, in mixed ordering
    scheme with different choices of kinematic spectators of the
    photon splitting, both charged primary leptons (top row) or
    only the primary lepton the splitting photon was reconstructed to
    have been emitted from (bottom row), and the way in which the
    starting scale of the evolution is chosen, probabilistically
    (left column) or by always choosing the winning dipole (right column).
  }
  \label{fig:results:systematicstdR}
\end{figure}

To further investigate the effects of our results on specific
algorithmic choices, Fig.\ \ref{fig:results:systematicstdR} focuses on 
the same observable familiar from the previous figure:
the interdependence of the starting scale
$\tstart$ and the opening angle $\Delta \Theta_\text{pair}$.
Here as in Fig.\ \ref{fig:results:startingscale}, we see that 
the effect of a winner-takes-all choice of starting scale as opposed 
to our default probabilistic starting scale definition is not 
significant. The winner-takes-all choice results in the distribution 
of starting scales being skewed to slightly smaller values, as discussed above,
which correlates loosely with a more collinear splitting. 
This results in a slight extension of the high-frequency (red) 
region of the plot towards the small-$\tstart$ small-angle corner 
in the lower two plots of Fig.\ \ref{fig:results:systematicstdR}
compared to the upper two plots. 
We also show the spectator
scheme dependence in the photon splitting: whether we allow both
primary leptons to be spectators or only the lepton that the photon
was reconstructed to have been emitted from.
Since in photon splittings, the spectator's only role is to absorb
recoil to guarantee momentum conservation, it is physically well
motivated for the splitting photon's progenitor to be the sole particle
to absorb its gained virtuality necessary for the splitting
process.
Note that this choice does not affect the value of $\tstart$, only the energy
available in the splitting, which affects the overall splitting probability
and the allowed opening angle of the splitting products. Fig.\ 
\ref{fig:results:systematicstdR} shows that this choice has negligible 
effect on the distribution of splitting events in the
$(\tstart,\Delta\Theta_\text{pair})$ plane.

\begin{figure}[t!]
  \centering
  \includegraphics[width=0.5\textwidth]{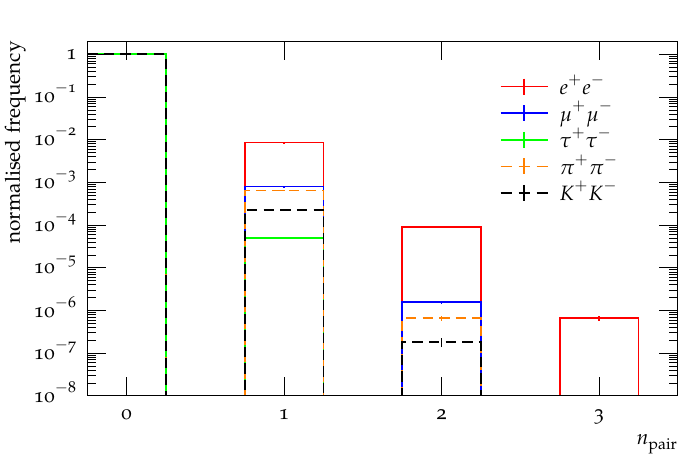}
  \caption{
    The relative abundance of secondary pairs of each species
    of charged particle produced in photon splittings in
    the mixed ordering scheme.
    \label{fig:results:number}
  }
\end{figure}

To conclude this section, Fig.\ \ref{fig:results:number} shows the
relative frequency of photons splitting into different species of
charged lepton and hadron.
As the driving factor is the produced particle species' mass,
electron-positron pairs are most commonly produced,
around an order of magnitude more commonly the products of
photon splittings than muons or charged pions.
The probability of producing a second pair of a given species
roughly follows the na\"ive expectation of being the square
of the probability of producing one pair.
In a more detailed consideration one finds a factor of
$\alpha^2\log(m_Z/E_\gamma)\,\log(\tstart/m^2)$ associated
with each secondary pair production.
Therein, $E_\gamma$ is the energy of the bremsstrahlungs photon
that subsequently splits into the pair of particles of mass $m$,
and $\tstart$ is its reconstructed starting scale.
Hence, we observe a single-logarithmic suppression
of heavier flavours, modulo possible minor differences in the splitting
function itself.
This is well-reproduced by our algorithm.
In fact, in the current example, the drop in frequency of
producing an additional pair of
particles of the same flavour is between 2.5 and 4.5 orders of magnitude.

\section{Lepton dressing beyond photons}
\label{sec:dress}

In this section we analyse the final states produced by our algorithm,
and in particular the consequences of further
resolving the photons produced by the standard soft-photon resummation
into charged-particle pairs.
We will continue to use the decay of an on-shell $Z$ boson
into an $e^+e^-$ pair as a testbed for our algorithm.
We will analyse the corrections induced by photon splittings 
on a number of physical properties that are related to the charged
particle content of the radiation cloud surrounding the primary
decay products.

The reader is reminded that we continue to use
a spherical coordinate system to measure relative radial
distances $\Delta\Theta$.
Further, please note that for this study we turn off kaon and $\tau$ decays
for the greatest accuracy in identifying primary final-state particles.
By default, however, kaon and $\tau$ lepton decays would be handled
as normal in \Sherpa \cite{Bothmann:2019yzt}, including various
state-of-the-art parametrisations of all known decay channels and
including their own respective QED corrections.

\subsection{Dressing strategies in the presence of photon splittings}
\label{sec:results:dress}

Lepton dressing is commonly used to define infrared-safe observables
through recombining a primary bare lepton with its surrounding
radiation cloud, in full analogy with the jet clustering of QCD.
While lepton dressing is essential when massless leptons are used
in a calculation due to the presence of collinear singularities,
the inclusion of a lepton mass renders both dressed and bare lepton
definitions physical.
Nonetheless, bare leptons suffer from large corrections that are
logarithmic in the lepton's mass, making them particularly
relevant for electrons.
Hence, a dressed lepton definition is also advantageous in calculations
with massive, but light, leptons.

In practice, there are two common methods for lepton dressing,
analogous to jet definitions in QCD: cone dressing and sequential
recombination dressing.
While a sequential recombination algorithm typically uses either
the anti-$k_t$ or Cambridge-Aachen algorithm \cite{Cacciari:2008gp},
the cone-based dressing uses the bare lepton to define the cone axis and,
at variance with historical QCD cone algorithms, keeps the cone
itself stable throughout the recombination procedure, rendering it
collinear safe.
In either case, the algorithm is not completely blind to particle flavour
since (at least) the primary bare lepton is used as the dressing-initiator
and defines the flavour of the resulting dressed lepton.
As long as only photon radiation is considered as a higher-order
correction to lepton production, which is the current standard in both
YFS based soft-photon resummations \cite{Hamilton:2006xz,Schonherr:2008av}
and \Photos \cite{Barberio:1993qi,Davidson:2010ew}, both algorithms work
very straightforwardly by subsequently combining the primary lepton
with the surrounding photon cloud using the respective distance
measure.

When photon splittings are included in the QED corrections to lepton
production as well, the radiation cloud surrounding the primary
lepton becomes flavour-diverse.
Considering the underlying physical process, these photon-splitting
corrections are simply resolving the structure of the photons
constituting the above photon cloud.
While these corrections are infrared finite when all lepton masses
are considered, large logarithmic effects can be expected in particular
when branching into the lightest species, electrons, occurs. Further,
the splitting into electrons is the most probable branching for a photon 
emitter.
Thus, while continuing to dress the primary leptons with photons only is
infrared safe, it is natural to demand that the resulting dressed lepton
definition does not strongly depend on whether or not we include further 
photon splittings.
We will thus investigate the following choices for the flavour set
$\fdress$ which is used to dress the primary lepton:
\begin{tabbing}
  \quad
  \=$\boldsymbol{\{\gamma\}}$\qquad\qquad\qquad\quad
  \=We continue to use only photons to dress the primary charged lepton.\\[1mm]
  \>$\boldsymbol{\{\gamma,e\}}$
  \>In addition to the mandatory dressing with the surrounding photons,
    we also\\
  \>\>
    include the lightest charged particle, the electron, in the
    dressing procedure of\\
  \>\>
    the primary lepton.
    This is not only motivated by the fact that splittings into\\
  \>\>
    $e^+e^-$
    pairs give the largest corrections, but also that experimentally
    both elec-\\
  \>\>
    trons and photons are measured similarly in the calorimeter.
    Of course, the\\
  \>\>
    presence of a magnetic field between the interaction
    and the calorimeter does\\
  \>\>
    in principle decorrelate the direction of
    their respective momentum vectors.\\[1mm]
  \>$\boldsymbol{\{\gamma,e,\pi,K\}}$
  \>We also include the lightest hadronic splitting products in the
    dressed lepton\\
  \>\>
    definition.
    Such a definition is a compromise between theoretical inclusivity
    and\\
  \>\>
    experimental feasability.\\[1mm]
  \>$\boldsymbol{\{\gamma,e,\pi,K,\mu,\tau\}}$
  \>We include all species produced in our photon-splitting implementation
    in \\
  \>\>
    order to be completely inclusive.
    It has to be noted though, that in realistic \\
  \>\>
    experimental environments
    muons are well distinguishable even at low muon \\
  \>\>
    energies, and $\tau$
    leptons of course decay further before detection rendering their\\
  \>\>
    inclusion in any realistic dressing algorithm highly non-trivial.
\end{tabbing}

\begin{figure}[t!]
  \centering
  \includegraphics[width=0.47\textwidth]{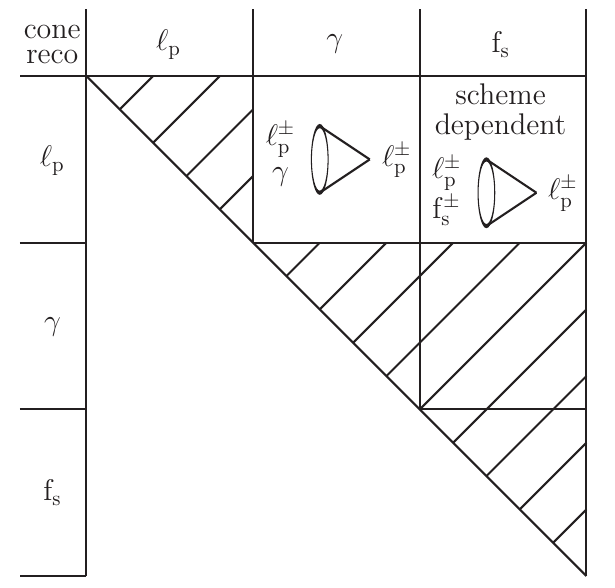}
  \hfill
  \includegraphics[width=0.47\textwidth]{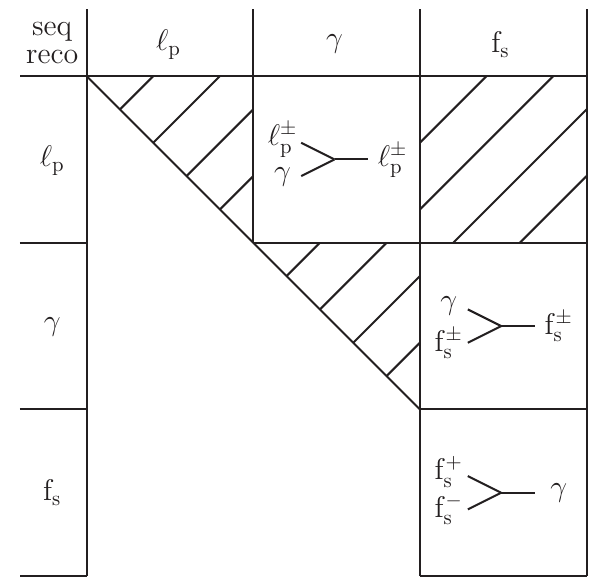}
  \caption{
    Recombination matrices of lepton dressing strategies beyond photon radiation.
    While $\gamma$ labels the photon, $\ellp$ and $\ells$ denote
    the primary leptons and secondary flavours, respectively.
    \label{fig:results:flavourmatrix}
  }
\end{figure}

A schematic of how both the cone and sequential recombination
dressing algorithms in the presence of photon splittings proceed
is given in Fig.\ \ref{fig:results:flavourmatrix}.
In the case of the case of cone dressing, the primary leptons
should be identified, and should be dressed with all QED radiation
that surrounds them, including other leptons and hadrons.
In particular, the flavour of the dressed lepton does not change
even if flavours other than a photon are included in it as it
is determined entirely by the primary lepton.
Thus, in consequence, the cone-dressed lepton may have a net
charge that is different from that of its assigned flavour when
not all photon-splitting products are recombined into the same
dressed lepton.
We will use this algorithm for the remainder of this study.

Nonetheless, a diagram of a flavour recombination matrix for sequential
recombination dressing is shown on the right-hand side of
Fig.\ \ref{fig:results:flavourmatrix}.
Here it is possible to recombine a secondary (and hence soft/collinear)
lepton-antilepton pair into a photon, while allowing for even
softer or more collinear surrounding photons to be combined with
these charged leptons first.
On the level of primary leptons, then, they are only dressed
with photons, either from the final state or from previous
secondary-lepton clusterings.
This has the obvious advantage that the charge and flavour
of the primary lepton matches that of the dressed lepton.
It, however, is schematically more intricate and does not
always lead circular dressed leptons, which are favoured
experimentally, and the investigation is left to a future study.

\begin{figure}[t!]
  \centering
  \includegraphics[width=0.47\textwidth]{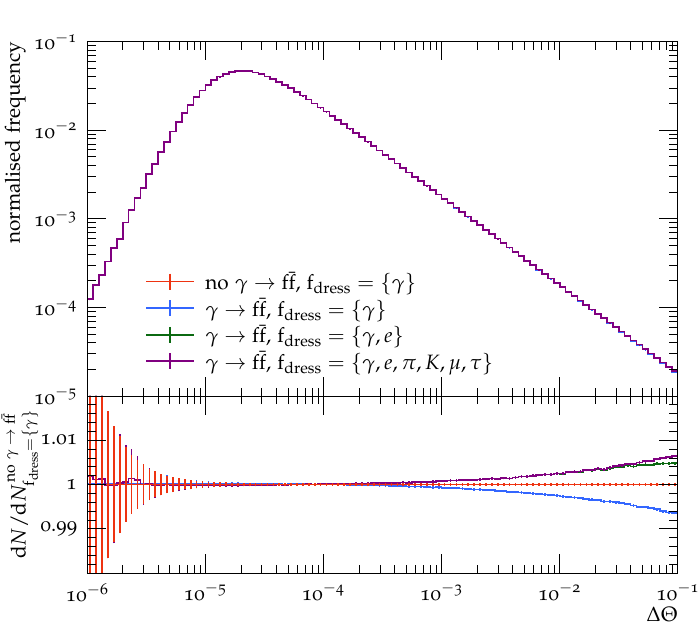}
  \hfill
  \includegraphics[width=0.47\textwidth]{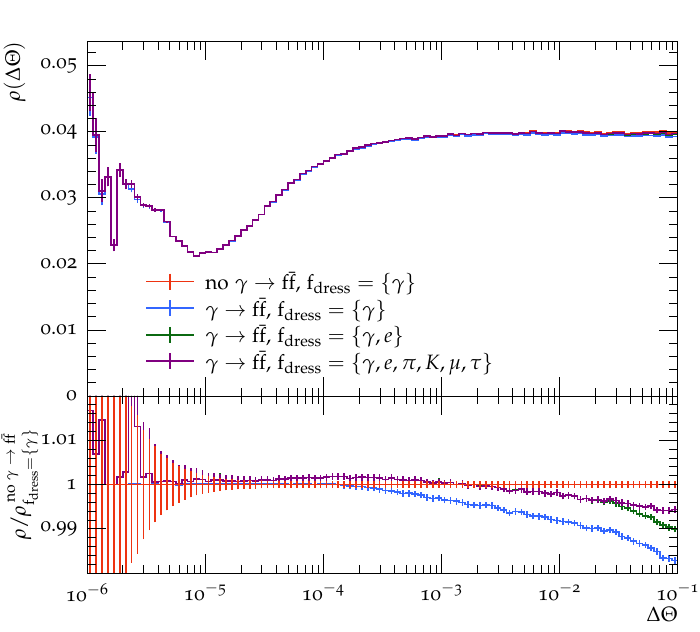}
  \caption{
    \textbf{Left:} The differential distribution of the dressed lepton
    constituents, including radiated photons with
    $E_\gamma>0.1\,\text{MeV}$, in dependence on the angular
    distance $\Delta\Theta$ from the primary lepton.
    \textbf{Right:} The energy density $\rho$ within the dressed lepton
    as a function of the angular distance $\Delta \Theta$ from the primary
    lepton.
    Shown are the predictions without accounting for photon splittings (red),
    compared to the predictions allowing photons to split: dressed
    with photons only (blue), photons and electrons (green) or all particles (violet).
    \label{fig:results:dtheta_rho}
  }
\end{figure}

The first observables we examine offer closer looks into the
substructure of the cone-dressed leptons produced by different
dressing strategies.
Here and in the following we use the notation:
either photon splittings to charged flavours $\mathrm{f}$ are
present ($\gamma \to \ffbar$) or they are
not (no $\gamma \to \ffbar$);
the dressing algorithm is specified by the set of particle
flavours \fdress\ which are included in the dressing.

The left-hand side plot of Fig.\ \ref{fig:results:dtheta_rho}
displays the angular distance $\Delta\Theta$ of the cone-dressed
lepton constituent from the primary lepton.
To ensure infrared safety, only photons with $E_\gamma>0.1\,\text{MeV}$
are included. A cutoff just below the electron mass has been selected 
to ensure that all electrons are included in the analysis.
Besides observing the primary lepton's dead cone for
$\Delta\Theta\lesssim 2\cdot 10^{-5}$,
we find that for $\Delta\Theta\lesssim 10^{-4}$ the constituent
multiplicity when including photon splittings, irrespective of the
dressing scheme used, coincides with the multiplicity when omitting
such splitting.
This corroborates our earlier expectation that collinear photons
largely lack the necessary virtuality to split into a charged-particle
pair.
At larger angles, where the required virtuality can be more easily
gained, a photon's probability to split increases.
In consequence, when including photon-splitting effects in the calculation,
but not accounting for the splitting products in the dressing, a drop
in multiplicity can be observed.
Including electrons as well as photons in the dressing reincorporates
most splitting products into the dressed lepton definition (\see Fig.\
\ref{fig:results:number}). We find an increase above the reference 
of approximately the same number of constituents that are lost in the 
$\gamma\to \ffbar$, $\fdress=\{\gamma\}$ (blue) case. 
The completely flavour-inclusive dressing definition then shows
the same effects scaled to the production of heavier secondary species:
larger virtualities, and thus larger $\Delta\Theta$, are needed for 
a non-zero splitting probability, so fewer photons actually split into
these heavier flavours. This leads to a much smaller effect of these 
splittings, concentrated at the outside of the cone. We expect out-of-cone 
effects to be small, since the frequency spectrum falls steeply towards 
the edge of the cone. More generally, it appears that the splitting products 
are close to collinear with the progenitor photon, at least on average. 

On the other hand, the right-hand side plot of
Fig.\ \ref{fig:results:dtheta_rho} shows the
distribution of energy within the dressed lepton as
a fraction of the energy of the entire dressed lepton.
Resolving photons into other species, \ie pairs of
charged particles, but continuing to dress the
primary lepton with photons only, naturally decreases
the energy radial density of the dressed lepton.
The fact that this energy density loss is not constant
but rather increases with the radial distance to the
primary lepton is again a result of the increasing
possible off-shellness at larger $\Delta\Theta$, and
therefore the increased splitting probability.
Even when the photon splittings products are part of
the dressing procedure, either secondary electrons
only or the set $\{e,\pi,K,\mu,\tau\}$, the energy density $\rho$
falls below the reference at some distance from the
primary lepton showing that a significant number of
splitting products end up outside the dressing cone radius.

\begin{figure}[t!]
  \centering
  \includegraphics[width=0.47\textwidth]{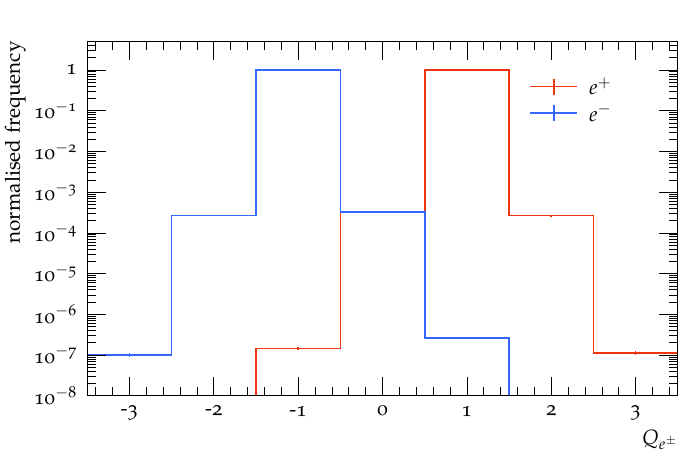}
  \caption{
    The total charge of the cone-dressed electron and positron with
    $\dRdress=0.1$ and including all secondary flavours,
    \ie $\fdress=\{\gamma,e,\pi,K,\mu,\tau\}$.
    \label{fig:results:charge}
  }
\end{figure}

As mentioned above, it is possible for the charge of the
dressed lepton to be different from the charge of its
primary constituent.
This is shown in Fig.\ \ref{fig:results:charge} for the
case of cone dressing with $\dRdress = 0.1$.
Fewer than a thousandth of the dressed leptons are neutral
or doubly charged, while a fraction of $10^{-7}$ of them
are either triply charged or appear to be their own
antiparticle (a dressed electron having a charge of $+1$
or a dressed positron having a charge of $-1$).
Again, this is a consequence of only partially capturing
the photon splitting products.

In the next section we will look at the separate and combined effects of photon splittings 
and flavour-aware lepton dressing on physical observables in the decay of an on-shell 
$Z$ boson.

\newpage
\subsection{Case study: \texorpdfstring{$Z$}{Z} boson decay}
\label{sec:results:DY}

In a final step, we look at the decay of an on-shell $Z$ boson into an
$e^+ e^-$ pair and investigate the impact of the photon splitting
corrections introduced in this paper on physical observables.
To be precise, we present the effects of including $\gamma \to \ffbar$
splittings and the consequences of (not) using flavour-aware dressing algorithms
on the decay rate, differential with respect to the invariant mass
$m_{\ell\ell}$ of the primary electron-positron pair.

\begin{figure}[t!]
  \centering
  \includegraphics[width=0.47\textwidth]{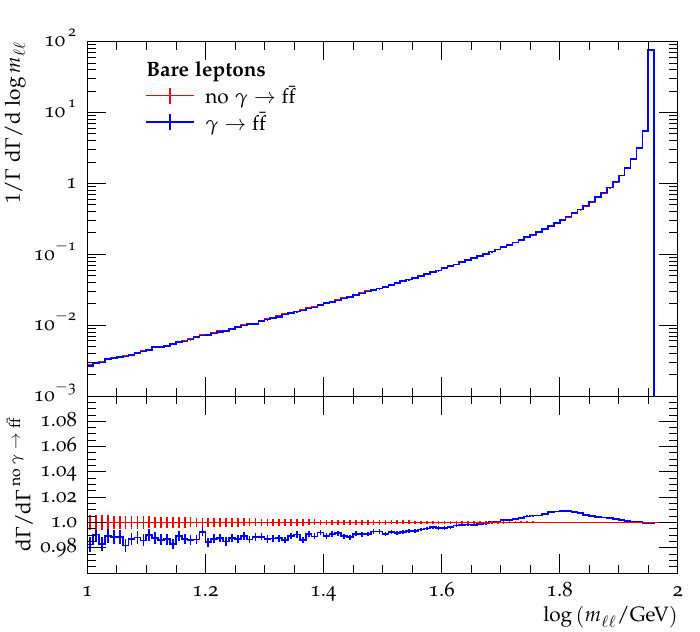}
  \caption{
    The bare dilepton invariant mass $m_{\ell\ell}$ as described by the
    YFS soft-photon resummation only (red) or additionally resolving the
    photons further into pairs of charged particles (blue), in the mixed
    ordering scheme.
    \label{fig:results:mllCompareBare}
  }
\end{figure}

We begin by examining the bare differential decay rate, \ie the invariant
mass of the primary lepton pair that is not dressed with the radiation
around it, in order to quantify the kinematic effect of photon splittings 
on the primary leptons themselves without confusing this effect with the 
intricacies of the dressing algorithm.
We note that bare leptons are theoretically well defined as all lepton masses
are fully accounted for.
To this end, Fig.\ \ref{fig:results:mllCompareBare} isolates the
effect of allowing YFS photons to split by presenting the bare
invariant mass of the two most energetic leptons of opposite charge,
one electron and one positron.
In the overwhelming majority of cases these are expected to be the
primary electron-positron pair generated in the on-shell $Z$ decay.
The largest deviation from the pure YFS prediction without photon splittings,
which is taken as the reference, is about $1\%$ in the region of
most interest.
It occurs just below the $Z$ mass, at about $60-70\,\GeV$.
It is driven by extracting additional momentum from the primary
leptons to accommodate the necessary virtuality for
photon splittings to occur.
Although barely visible, this is fueled by a minute reduction
of the much larger differential decay rate closer to the
nominal $Z$ mass itself.
Although of less interest due to the smaller absolute decay rate,
the opposite effect is seen at very small invariant masses, below
$50\,\GeV$, where, through the same mechanism, the decay rate is
diminished by about $1-2\%$ as the slope of the distribution is
shallower but the momentum extraction is similar in magnitude to that 
at larger invariant masses. 

Finally, a change of the precise definition of the ordering
variable, both for the reconstructed starting scale of the evolution
and the splitting scale of the eventual photon splitting, generally
increases the size of the corrections for this observable.
While using $t=\kTsq$ for all splittings only increases the observed
corrections slightly, due to the increased photon splitting probability
as $\kT<\qbar$ throughout, using $t=\qbar^2$ almost doubles the size
of the corrections as now the starting scales of the each photon's
evolution reconstruct to much larger values, \see
Fig.\ \ref{fig:results:startingscale}.
This is a consequence of the different properties of these 
ordering variables as discussed in section \ref{sec:methods:photonsplit},
although \textit{a priori} all choices have the same formal accuracy.

\begin{figure}[t!]
  \centering
  \includegraphics[width=0.47\textwidth]{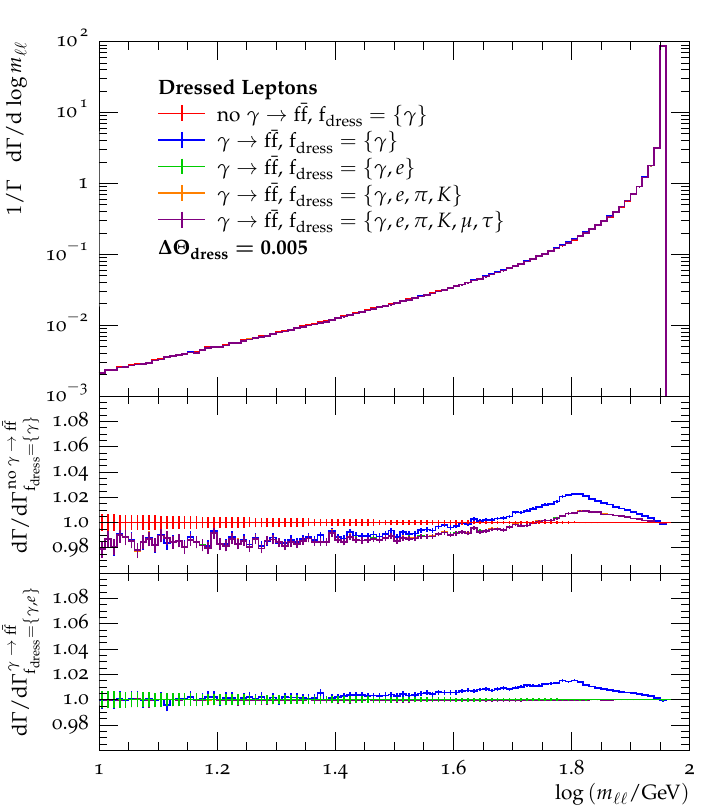}
  \hfill
  \includegraphics[width=0.47\textwidth]{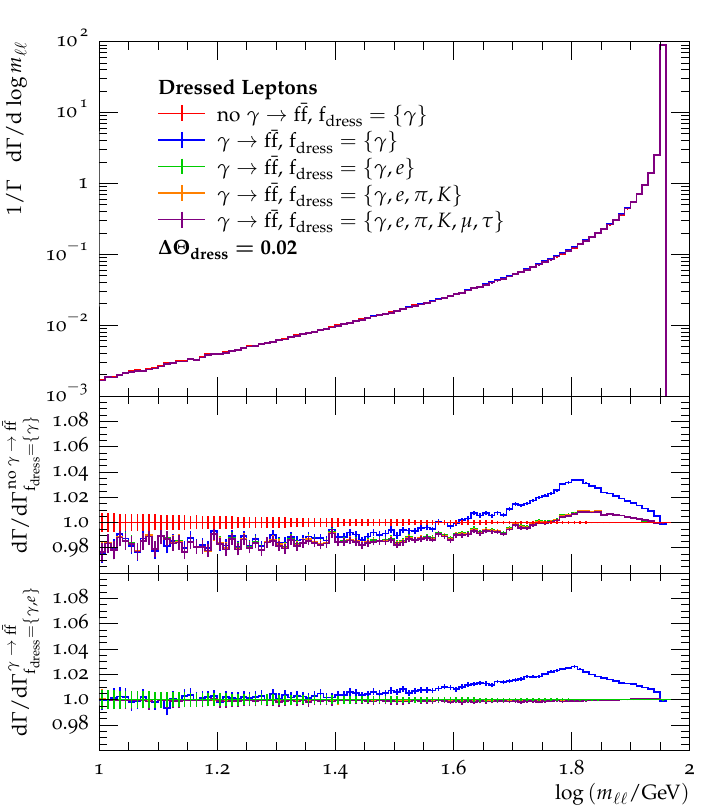}

  \includegraphics[width=0.47\textwidth]{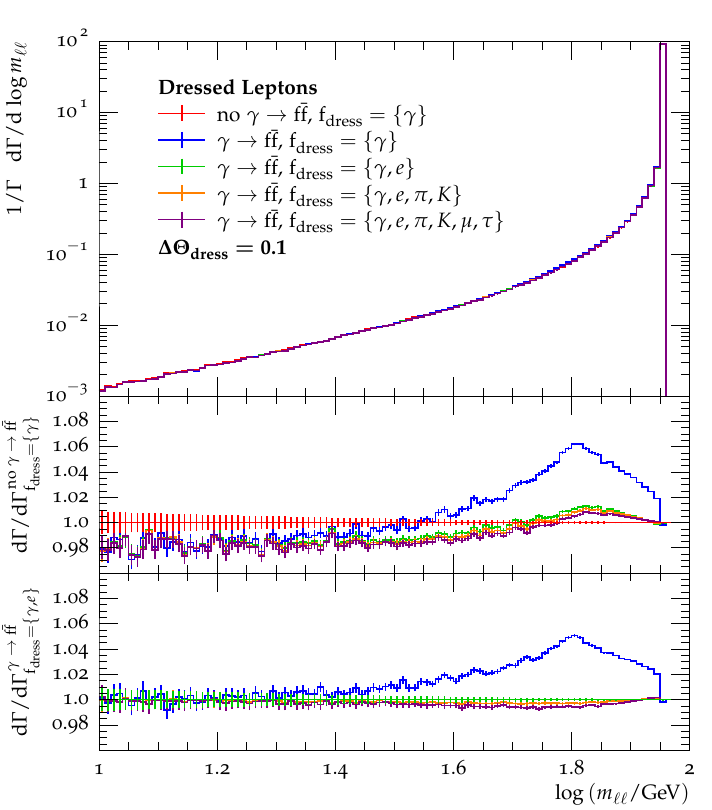}
  \hfill
  \includegraphics[width=0.47\textwidth]{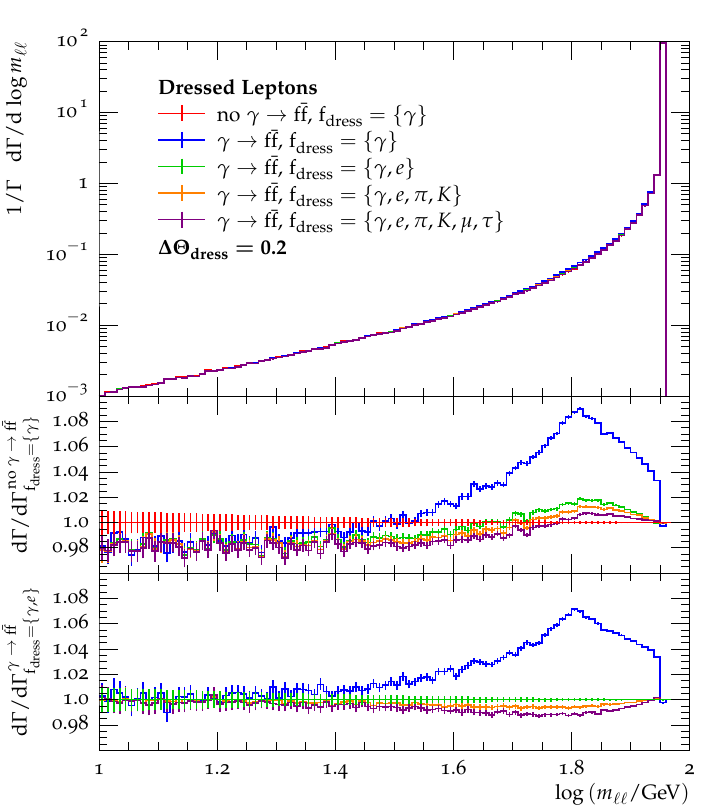}
  \caption{
    The dressed dilepton invariant mass $m_{\ell\ell}$ as described by the
    YFS soft-photon resummation only (red) or additionally resolving the
    photons further into pairs of charged particles for four different
    dressing cone sizes, $\dRdress=0.005$ (top left), $0.02$ (top right),
    0.1 (bottom left), and 0.2 (bottom right), in the mixed ordering scheme.
    We differentiate various different dressing strategies, recombining
    photons only (blue), photons and electrons (green), photons,
    electrons and charged hadrons (orange), and all charged particles
    (violet) within the dressing cone with the primary charged lepton.
    Two ratios are presented, either taking the soft-photon resummation
    without photon splittings (upper ratio), or the soft-photon resummation
    including photon splittings and dressing the primary leptons with
    photons as well as secondary electrons (lower ratio),
    as the reference.
    \label{fig:results:mllCompare3}
  }
\end{figure}

Having assessed the basic kinematic effects on the bare primary
leptons, we now turn to dressed leptons.
We will investigate the impact the different dressing strategies
discussed in Sec.\ \ref{sec:results:dress} have once the radiation cloud around
the primary leptons is not comprised of only photons but is resolved
further into various different flavours of secondary charged particles.
To this end, Fig.\ \ref{fig:results:mllCompare3} contrasts the pure YFS
soft-photon resummation without further photon splittings with a range of
dressing strategies when photon splittings are included.
Four different cone sizes are considered, from $\dRdress = 0.005$ to
$\dRdress = 0.2$.
The upper ratio illustrates the deviation of each prediction
from the pure YFS case, due to both the presence of photon splittings
and the details of the dressing algorithm.
The lower ratio isolates the effect of the dressing strategy
by showing the deviation with respect
to the photon-only-dressed events. In particular, this shows 
which secondary flavours are recombined with the primary
lepton into the dressed lepton.
We observe that when the photon radiation off the primary electrons
is further resolved into charged-particle pairs but the primary
electrons are still only dressed with only the photons of their
surrounding radiation cloud, large effects are manifest.
They range from slightly over 2\% for $\dRdress = 0.005$ to 6\%
for the common $\dRdress = 0.1$, and up to 9\% for the more inclusive
cone radius of $\dRdress = 0.2$.
This difference originates in the fact that as long as only photons
are included in the dressing, every photon lost by resolving it into
a charged-particle pair cannot be recombined into the dressed lepton,
which then ends up with less energy simply because higher-order
corrections have been included.
The observation that our algorithm reconstructs higher starting
scales for hard wide-angle photons
than either soft or collinear ones, and thus these are 
more likely to possess the necessary virtuality to split into
charged-particle pairs, explains the dressing-cone-size dependence.
However, when more inclusive dressing algorithms are considered,
the effect of photon splittings on the differential decay rate is
reduced, as is the $\dRdress$ dependence.
As photons predominantly resolve into $e^+e^-$ pairs, their inclusion
in the dressed lepton definition already captures the bulk of the
effect, in particular at smaller dressing cone radii.
Along the lines of the above argument, photons need to be sufficiently separated
from the primary lepton in order to gain enough
virtuality to split into the heavier particle species.
Thus, the inclusion of further secondary flavours in the dressing
algorithm only plays a role at larger dressing cones, with effects
ranging from 1\% at $\dRdress=0.1$ to 2\% at $\dRdress=0.2$.
The effect of changing the ordering scheme for the photon splitting 
algorithm on Fig.\ \ref{fig:results:mllCompare3} is very similar to the effect
on Fig.\ \ref{fig:results:mllCompareBare}. Again, using the transverse momentum
or virtuality ordered schemes increases the size of 
the corrections induced by photon splittings in a very similar way as before.
It is still the case that reincorporating splitting products in the dressing
recovers the bare-lepton level deviation from the pure YFS prediction.
As above, it needs to be noted that such a change in the ordering
variable results in a subobtimal description of the physical process,
and is thus not recommended to be used as an estimator of the intrinsic
uncertainty.

\begin{figure}[t!]
  \centering
  \includegraphics[width=0.47\textwidth]{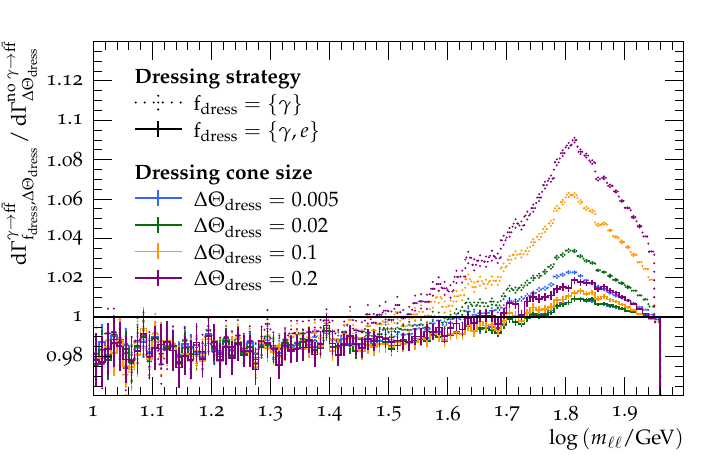}
  \hfill
  \includegraphics[width=0.47\textwidth]{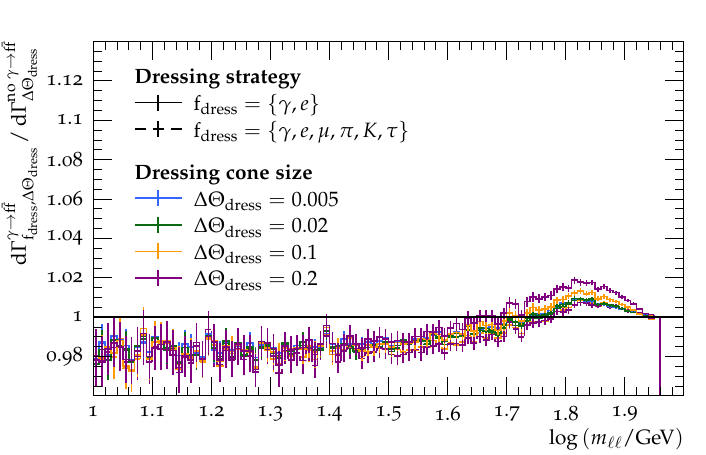}
  \caption{
    This figure shows the cone size dependence of different dressing
    strategies. The differential decay rate
    $\mathrm{d}\Gamma^{\gamma \to \ffbar}_{\fdress,\dRdress}/
    \mathrm{d}\log{m_{\ell \ell}}$
    has been divided by the corresponding
    $\mathrm{d}\Gamma^{\mathrm{no}\,\gamma \to \ffbar}_{\dRdress}/
    \mathrm{d}\log{m_{\ell \ell}}$,
    in dependence of both the flavour set $\fdress$ included
    in the dressing and the dressing cone of size $\dRdress$.
    The \textbf{left} plot shows the difference case where only
    photons are used in the dressing (dotted) and using both
    photons and secondary electrons (solid),
    whereas the \textbf{right} plot shows the difference between
    a dressing strategy using only photons and electrons (solid)
    and all secondary flavours (dashed).
    \label{fig:results:mllDivided}
  }
\end{figure}

In Fig.\ \ref{fig:results:mllDivided} we show more clearly the recovery 
of the pure soft-photon prediction using the two most relevant 
charged-particle-inclusive dressing strategies.
The figure shows the ratio of the differential
cross section including photon splittings to that without photon splittings 
for different dressing choices.
We find that
including charged particles in the cone dressing limits the effect of
photons splitting corrections to the 1\% level, irrespective of cone size.
Including electrons in the dressing similarly limits the corrections 
to 2\% even for the largest cone sizes considered here.

\section{Conclusions}
\label{sec:conclusions}

In this paper we detailed an extension to the soft-photon resummation
in the Yennie-Frautschi-Suura framework to incorporate higher QED
corrections originating in photon splittings into charged-particle pairs.
These photon-splitting corrections, which resolve the substructure of 
the newly produced photons, are often larger than suggested by 
the formal accuracy. In particular, they can be 
logarithmically enhanced with the ratio of the lightest charged 
particle, the electron, to the possible virtuality of the 
splitting photon. 

Using the decay $Z\to e^+e^-$, we found that that the limit on
the virtuality of the photon bremsstrahlung off a primary lepton 
is strongly correlated with the angular distance to this primary lepton, and
thus also to the probability of that photon to split.
We also investigated the systematics of our photon-splitting algorithm 
and found that algorithmic choices do not have a large
impact on results. 
We found that the frequency of occurrence of different species 
agreed with theoretical expectations, showing a logarithmic dependence 
on the mass of the produced particles.

As a consequence of our extension, the cloud of QED radiation 
surrounding the primary leptons of a hard decay contains an 
array of particle flavours, not solely photons.
Hence, the standard dressing algorithms
to define infrared-safe dressed leptons were found to develop
a strong sensitivity to further resolving the initial soft-photon
cloud, in particular for larger dressing-cone radii.
We therefore developed a novel set of flavour-aware strategies
for dressing charged leptons and investigated their respective
properties.
We found that including secondary electrons as a minimal addition
in the dressing procedure already substantially reduces this
dependence on photon resolution, while an inclusion of all possible 
secondary flavours minimises it.

Using the example of the $Z\to e^+e^-$ decay rate, we investigated
the dilepton invariant mass in detail.
We found corrections of around 1\% from photon splittings, wrt.\
the previous standard of not further resolving the initial photon
radiation on the bare electrons.
In the more relevant case of leptons cone-dressed with photons only,
these could become much larger, up to 9\% for large dressing
cone radii of $\dRdress=0.1$ or $0.2$.
Introducing a flavour-aware dressing algorithm restored the
bare result to a large degree, however, reigning in the
photon-splitting corrections to $1-2\%$, along with mostly
removing their cone-size dependence.
We leave it to a future publication to study how the above effects
translate to the general off-shell production of a Drell-Yan
lepton pair at a hadron collider.

The photon splitting corrections were implemented in the
\Sherpa Monte-Carlo event generator and will be incorporated in a future
release.
All analyses and dressing strategies were implemented
using \Rivet's analysis tools \cite{Buckley:2010ar,Bierlich:2019rhm}.

\subsection*{Acknowledgements}

LF would like to thank Hitham Hassan for valuable discussions on this topic.
MS is funded by the Royal Society through a University Research Fellowship
(URF\textbackslash{}R1\textbackslash{}180549) and Enhancement Awards
(RF\textbackslash{}ERE\textbackslash{}210397,
 RGF\textbackslash{}EA\textbackslash{}181033 and
 CEC19\textbackslash{}100349).
LF is supported by the UK Science and Technology Facilities
Council under contract
ST/T001011/1.

\appendix
\section{Charged resonances}
\label{app:fidip}

In this appendix we give the definitions for final-initial (FI) dipoles needed
for the description of photon splittings in the QED corrections
of charged particle decays, like $W\to\ell\nu$.

The notation used in this appendix is for the most part consistent with Sec.\
\ref{sec:methods:photonsplit}. We consider a charged resonance $\at$ ($a$)
decaying to a charged particle $\ijt$ ($i$) and a recoiling system
$\{\tilde{n}\}$ ($\{n\}$):\, $\at\to\ijt\,\{\tilde{n}\}$ before and
$a\to i\,j\,\{n\}$ after the emission of a photon $j$, respectively.
Ordinarily the recoil from the splitting would be absorbed locally by
either the spectator $a$ when $i$ is the emitter, or vice versa.
Since $a$ is the decaying particle, however, we have chosen
to keep its momentum unchanged and redistribute the recoil effectively to the
particle(s) $\{n\}$.
This allows to use a single momentum map for both situations and
combine both emitter-spectator designations into a single dipole
splitting function.
This not only simplifies its description, but also removes problems
with the positivity of the partial-fractioned
Catani-Seymour splitting functions in situations where the mass correction
is larger than the (quasi-)collinear emission term.
Hence, we follow the treament in \cite{Basso:2015gca,
  Dittmaier:1999mb} to construct the splitting functions and kinematic
variables.

As described in Sec.\ \ref{sec:methods}, the first step in the photon
splitting algorithm is to determine the starting scale
of each photon by reconstructing its emission history.
In principle, emission of a photon can occur from the decaying particle $\at$
or from its charged decay product $\ijt$. However, since the former splitting is 
suppressed by the decaying particle's mass, it is much more likely to act
as spectator.
Instead, as discussed above, we employ a single splitting
function which contains the initial-state emission term in addition to 
the final-state emission.
As a consequence, in the soft limit the full eikonal is recovered
and the dipole radiates coherently,
but splitting from the initial-state particle is never kinematically 
considered when building the required single-emitter history in the
collinear interpretation of our parton shower.

After calculating the starting scale, the photons' evolution begins and 
photon splittings are considered. Since a photon is never considered to be 
emitted from the initial-state charged particle, and the decaying particle 
has a restricted phase space for absorbing recoil in any case, the spectator 
in all photon splittings is chosen to be the final-state particle $i$.
For this reason, we do not need the kinematic mappings for an FI dipole.
The splitting function and evolution variable definitions are detailed 
below.

\paragraph*{Splitting functions.}

The dipole invariant mass $Q^2$ is defined as 
\begin{equation}
  Q^2 = (\tilde{p}_{\ijt}-\tilde{p}_{\at})^2 = (p_i+p_j-p_a)^2
\end{equation}
for the case of a dipole with final-state emitter $i$ and initial-state 
spectator $a$.

For convenience, we define the quantity 
\begin{equation}
  \bar{Q}^2 = m_a^2 - m_i^2 -m_j^2 - Q^2.
\end{equation}

The kinematic variables $z$ and $y$ are defined differently from the FF
case; they are given by 
\begin{equation}
  y = \frac{p_i p_j}{p_a p_i + p_a p_j - p_i p_j -2m_i^2 -2m_j^2}
  \qquad\text{and}\qquad
  z = \frac{p_i p_a - p_i p_j -m_i^2}{p_i p_a + p_j p_a -2p_i p_j -m_i^2-m_j^2}.
\end{equation}

In terms of these variables the splitting functions are given by 
\begin{equation}
  \begin{split}
    S_{f_\ijt(\at)\to f_i\gamma_j(a)}
    \,=\;
    S_{\bar{f}_\ijt(\at)\to \bar{f}_i\gamma_j(a)}
    \,=&\;
      -\,\mathbf{Q}^2_{\ijt\at}\;\alpha\,
      \left[ \frac{2}{1-z(1-y)}
      \left(1+\frac{2m_i^2}{\bar{Q}^2}\right)
      -(1+z)-\frac{m_i^2}{p_i p_j} \right. \\
    &\;\hspace*{17mm}{}
      - \left.\frac{(p_i p_j)}{\bar{Q}^2} \,\frac{m_a^2}{\bar{Q}^2}\,
      \frac{4}{[1-z(1-y)]^2} \right], \\
    S_{s_\ijt(\at)\to s_i\gamma_j(a)}
    \,=\;
    S_{\bar{s}_\ijt(\at)\to \bar{s}_i\gamma_j(a)}
    \,=&\;
      -\,\mathbf{Q}^2_{\ijt\at}\;\alpha\,
      \left[ \frac{2}{1-z(1-y)}
      \left(1+\frac{2m_i^2}{\bar{Q}^2}\right)
      -2-\frac{m_i^2}{p_i p_j} \right. \\
    &\;\hspace*{17mm}{}
      - \left.\frac{(p_i p_j)}{\bar{Q}^2} \,\frac{m_a^2}{\bar{Q}^2}\,
      \frac{4}{[1-z(1-y)]^2} \right]. \\
  \end{split}
\end{equation}
The additional factor $(1+2m_i^2/\bar{Q}^2)$ is needed to recover the soft 
eikonal limit by cancelling some of the mass dependence of the variables 
$z$ and $y$. Note that $m_j=0$ needs to be taken for the soft limit so is 
not present in this additional factor.

\paragraph*{Evolution variable.} As before, we consider two choices of 
evolution variable, virtuality and transverse momentum. The form of these 
variables in terms of the dipole invariant mass $Q^2$ and the masses of 
the particles in the process are very similar to those for FF dipoles. 

The virtuality is given by 
\begin{equation}
  \qbar^2 = (m_a^2-m_i^2-m_j^2-Q^2)\,y+m_i^2+m_j^2-m_{\ijt}^2
\end{equation}
while the transverse momentum can be written 
\begin{equation}
  \kTsq = (m_a^2-m_i^2-m_j^2-Q^2) \,y\, z(1-z) - m_i^2\,(1-z)^2 - m_j^2\,z^2.
\end{equation}
As before, the default scheme for the evolution variable is the mixed 
scheme, where the transverse momentum is computed as the starting scale 
for photon evolution but is interpreted as a virtuality thereafter. The pure 
transverse momentum and virtuality schemes are implemented as well.

\section{Usage}
\label{app:settings}

In this appendix we list the available keywords and settings
in order to effect the various algorithmic choices described
in this paper.
They are
\begin{description}
  \item[\texttt{YFS\_PHOTON\_SPLITTER\_MODE}] \
    This setting governs which secondary flavours will be considered.
    \begin{itemize}
      \item[\texttt{0}] photons do not split,
      \item[\texttt{1}] photons split into electron-positron pairs,
      \item[\texttt{2}] muons,
      \item[\texttt{4}] tau leptons,
      \item[\texttt{8}] and/or light hadrons
                        (up to \texttt{YFS\_PHOTON\_SPLITTER\_MAX\_HADMASS}).
    \end{itemize}
    The settings are additive, the default is \texttt{15}.
  \item[\texttt{YFS\_PHOTON\_SPLITTER\_MAX\_HADMASS}] \
    This setting sets the mass of the heaviest hadron which can be
    produced in photon splittings.
    Note that vector splitting functions are not currently implemented.
    Default is $0.5\,\GeV$.
  \item[\texttt{YFS\_PHOTON\_SPLITTER\_ORDERING\_SCHEME}] \
    This setting defines the ordering scheme used.
    \begin{itemize}
      \item[\texttt{0}] transverse momentum ordering,
      \item[\texttt{1}] virtuality ordering,
      \item[\texttt{2}] mixed scheme (default).
    \end{itemize}
  \item[\texttt{YFS\_PHOTON\_SPLITTER\_SPECTATOR\_SCHEME}] \
    This setting defines the allowed spectators for the photon splitting
    process.
    \begin{itemize}
      \item[\texttt{0}] all primary emitters may act as spectator (default),
      \item[\texttt{1}] only the photon's reconstructed emitter is eligible
                        as a spectator.
    \end{itemize}
  \item[\texttt{YFS\_PHOTON\_SPLITTER\_STARTING\_SCALE\_SCHEME}] \
    This setting governs the determination of the starting scale.
    \begin{itemize}
      \item[\texttt{0}] starting scale is chosed probabilistically (default),
      \item[\texttt{1}] the starting scale is chosen using a
                        winner-takes-all strategy.
    \end{itemize}
\end{description}

\bibliographystyle{amsunsrt_modpp}
\bibliography{journal}

\ifx\mcitethebibliography\mciteundefinedmacro
\PackageError{amsunsrt_mod.bst}{mciteplus.sty has not been loaded}
{This bibstyle requires the use of the mciteplus package.}\fi
\begin{mcitethebibliography}{10}

\bibitem{CDF:2022hxs}
T.~Aaltonen et~al., CDF, \emph{{High-precision measurement of the W boson mass
  with the CDF II detector}}, Science \textbf{376} (2022), no.~6589,
  \href{http://www.slac.stanford.edu/spires/find/hep/www?j=Science,376,170}{170--176},
  FERMILAB-PUB-22-254-PPD%
\relax\mciteBstWouldAddEndPuncttrue
\mciteSetBstMidEndSepPunct{\mcitedefaultmidpunct}
{\mcitedefaultendpunct}{\mcitedefaultseppunct}\relax
\EndOfBibitem
\bibitem{ParticleDataGroup:2022pth}
R.~L. Workman et~al., Particle Data Group, \emph{{Review of Particle Physics}},
  PTEP \textbf{2022} (2022),
  \href{http://www.slac.stanford.edu/spires/find/hep/www?j=PTEP,2022,083C01}{083C01}%
\relax\mciteBstWouldAddEndPuncttrue
\mciteSetBstMidEndSepPunct{\mcitedefaultmidpunct}
{\mcitedefaultendpunct}{\mcitedefaultseppunct}\relax
\EndOfBibitem
\bibitem{LHCb:2021bjt}
R.~Aaij et~al., LHCb, \emph{{Measurement of the W boson mass}}, JHEP
  \textbf{01} (2022),
  \href{http://www.slac.stanford.edu/spires/find/hep/www?eprint=2109.01113}{036},
   [\href{http://arXiv.org/pdf/2109.01113}{{\tt arXiv:2109.01113}} [hep-ex]]%
\relax\mciteBstWouldAddEndPuncttrue
\mciteSetBstMidEndSepPunct{\mcitedefaultmidpunct}
{\mcitedefaultendpunct}{\mcitedefaultseppunct}\relax
\EndOfBibitem
\bibitem{ATLAS:2017rzl}
M.~Aaboud et~al., ATLAS, \emph{{Measurement of the $W$-boson mass in pp
  collisions at $\sqrt{s}=7$ TeV with the ATLAS detector}}, Eur. Phys. J. C
  \textbf{78} (2018), no.~2,
  \href{http://www.slac.stanford.edu/spires/find/hep/www?eprint=1701.07240}{110},
   [\href{http://arXiv.org/pdf/1701.07240}{{\tt arXiv:1701.07240}} [hep-ex]],
  [Erratum: Eur.Phys.J.C 78, 898 (2018)]%
\relax\mciteBstWouldAddEndPuncttrue
\mciteSetBstMidEndSepPunct{\mcitedefaultmidpunct}
{\mcitedefaultendpunct}{\mcitedefaultseppunct}\relax
\EndOfBibitem
\bibitem{CDF:2013dpa}
T.~A. Aaltonen et~al., CDF, D0, \emph{{Combination of CDF and D0 $W$-Boson Mass
  Measurements}}, Phys. Rev. D \textbf{88} (2013), no.~5,
  \href{http://www.slac.stanford.edu/spires/find/hep/www?eprint=1307.7627}{052018},
   [\href{http://arXiv.org/pdf/1307.7627}{{\tt arXiv:1307.7627}} [hep-ex]]%
\relax\mciteBstWouldAddEndPuncttrue
\mciteSetBstMidEndSepPunct{\mcitedefaultmidpunct}
{\mcitedefaultendpunct}{\mcitedefaultseppunct}\relax
\EndOfBibitem
\bibitem{CDF:2013bqv}
T.~A. Aaltonen et~al., CDF, \emph{{Precise Measurement of the W -Boson Mass
  with the Collider Detector at Fermilab}}, Phys. Rev. D \textbf{89} (2014),
  no.~7,
  \href{http://www.slac.stanford.edu/spires/find/hep/www?eprint=1311.0894}{072003},
   [\href{http://arXiv.org/pdf/1311.0894}{{\tt arXiv:1311.0894}} [hep-ex]]%
\relax\mciteBstWouldAddEndPuncttrue
\mciteSetBstMidEndSepPunct{\mcitedefaultmidpunct}
{\mcitedefaultendpunct}{\mcitedefaultseppunct}\relax
\EndOfBibitem
\bibitem{D0:2012kms}
V.~M. Abazov et~al., D0, \emph{{Measurement of the W Boson Mass with the D0
  Detector}}, Phys. Rev. Lett. \textbf{108} (2012),
  \href{http://www.slac.stanford.edu/spires/find/hep/www?eprint=1203.0293}{151804},
   [\href{http://arXiv.org/pdf/1203.0293}{{\tt arXiv:1203.0293}} [hep-ex]]%
\relax\mciteBstWouldAddEndPuncttrue
\mciteSetBstMidEndSepPunct{\mcitedefaultmidpunct}
{\mcitedefaultendpunct}{\mcitedefaultseppunct}\relax
\EndOfBibitem
\bibitem{CDF:2012gpf}
T.~Aaltonen et~al., CDF, \emph{{Precise measurement of the $W$-boson mass with
  the CDF II detector}}, Phys. Rev. Lett. \textbf{108} (2012),
  \href{http://www.slac.stanford.edu/spires/find/hep/www?eprint=1203.0275}{151803},
   [\href{http://arXiv.org/pdf/1203.0275}{{\tt arXiv:1203.0275}} [hep-ex]]%
\relax\mciteBstWouldAddEndPuncttrue
\mciteSetBstMidEndSepPunct{\mcitedefaultmidpunct}
{\mcitedefaultendpunct}{\mcitedefaultseppunct}\relax
\EndOfBibitem
\bibitem{D0:2009yxq}
V.~M. Abazov et~al., D0, \emph{{Measurement of the W boson mass}}, Phys. Rev.
  Lett. \textbf{103} (2009),
  \href{http://www.slac.stanford.edu/spires/find/hep/www?eprint=0908.0766}{141801},
   [\href{http://arXiv.org/pdf/0908.0766}{{\tt arXiv:0908.0766}} [hep-ex]]%
\relax\mciteBstWouldAddEndPuncttrue
\mciteSetBstMidEndSepPunct{\mcitedefaultmidpunct}
{\mcitedefaultendpunct}{\mcitedefaultseppunct}\relax
\EndOfBibitem
\bibitem{CDF:2007mxw}
T.~Aaltonen et~al., CDF, \emph{{First Run II Measurement of the $W$ Boson
  Mass}}, Phys. Rev. D \textbf{77} (2008),
  \href{http://www.slac.stanford.edu/spires/find/hep/www?eprint=0708.3642}{112001},
   [\href{http://arXiv.org/pdf/0708.3642}{{\tt arXiv:0708.3642}} [hep-ex]]%
\relax\mciteBstWouldAddEndPuncttrue
\mciteSetBstMidEndSepPunct{\mcitedefaultmidpunct}
{\mcitedefaultendpunct}{\mcitedefaultseppunct}\relax
\EndOfBibitem
\bibitem{CDF:2007cmy}
T.~Aaltonen et~al., CDF, \emph{{First measurement of the $W$ boson mass in Run
  II of the Tevatron}}, Phys. Rev. Lett. \textbf{99} (2007),
  \href{http://www.slac.stanford.edu/spires/find/hep/www?eprint=0707.0085}{151801},
   [\href{http://arXiv.org/pdf/0707.0085}{{\tt arXiv:0707.0085}} [hep-ex]]%
\relax\mciteBstWouldAddEndPuncttrue
\mciteSetBstMidEndSepPunct{\mcitedefaultmidpunct}
{\mcitedefaultendpunct}{\mcitedefaultseppunct}\relax
\EndOfBibitem
\bibitem{CDF:2003tdi}
V.~M. Abazov et~al., CDF, D0, \emph{{Combination of CDF and D0 Results on $W$
  Boson Mass and Width}}, Phys. Rev. D \textbf{70} (2004),
  \href{http://www.slac.stanford.edu/spires/find/hep/www?eprint=hep-ex/0311039}{092008},
   [\href{http://arXiv.org/pdf/hep-ex/0311039}{{\tt hep-ex/0311039}}]%
\relax\mciteBstWouldAddEndPuncttrue
\mciteSetBstMidEndSepPunct{\mcitedefaultmidpunct}
{\mcitedefaultendpunct}{\mcitedefaultseppunct}\relax
\EndOfBibitem
\bibitem{CDF:2000gwd}
T.~Affolder et~al., CDF, \emph{{Measurement of the $W$ boson mass with the
  Collider Detector at Fermilab}}, Phys. Rev. D \textbf{64} (2001),
  \href{http://www.slac.stanford.edu/spires/find/hep/www?eprint=hep-ex/0007044}{052001},
   [\href{http://arXiv.org/pdf/hep-ex/0007044}{{\tt hep-ex/0007044}}]%
\relax\mciteBstWouldAddEndPuncttrue
\mciteSetBstMidEndSepPunct{\mcitedefaultmidpunct}
{\mcitedefaultendpunct}{\mcitedefaultseppunct}\relax
\EndOfBibitem
\bibitem{D0:2002fhu}
V.~M. Abazov et~al., D0, \emph{{Improved $W$ Boson Mass Measurement with the D0
  Detector}}, Phys. Rev. D \textbf{66} (2002),
  \href{http://www.slac.stanford.edu/spires/find/hep/www?eprint=hep-ex/0204014}{012001},
   [\href{http://arXiv.org/pdf/hep-ex/0204014}{{\tt hep-ex/0204014}}]%
\relax\mciteBstWouldAddEndPuncttrue
\mciteSetBstMidEndSepPunct{\mcitedefaultmidpunct}
{\mcitedefaultendpunct}{\mcitedefaultseppunct}\relax
\EndOfBibitem
\bibitem{D0:1999rsi}
B.~Abbott et~al., D0, \emph{{A measurement of the $W$ boson mass using
  electrons at large rapidities}}, Phys. Rev. Lett. \textbf{84} (2000),
  \href{http://www.slac.stanford.edu/spires/find/hep/www?eprint=hep-ex/9909030}{222--227},
   [\href{http://arXiv.org/pdf/hep-ex/9909030}{{\tt hep-ex/9909030}}]%
\relax\mciteBstWouldAddEndPuncttrue
\mciteSetBstMidEndSepPunct{\mcitedefaultmidpunct}
{\mcitedefaultendpunct}{\mcitedefaultseppunct}\relax
\EndOfBibitem
\bibitem{D0:1999ivw}
B.~Abbott et~al., D0, \emph{{A measurement of the $W$ boson mass using large
  rapidity electrons}}, Phys. Rev. D \textbf{62} (2000),
  \href{http://www.slac.stanford.edu/spires/find/hep/www?eprint=hep-ex/9908057}{092006},
   [\href{http://arXiv.org/pdf/hep-ex/9908057}{{\tt hep-ex/9908057}}]%
\relax\mciteBstWouldAddEndPuncttrue
\mciteSetBstMidEndSepPunct{\mcitedefaultmidpunct}
{\mcitedefaultendpunct}{\mcitedefaultseppunct}\relax
\EndOfBibitem
\bibitem{DELPHI:2008avl}
J.~Abdallah et~al., DELPHI, \emph{{Measurement of the Mass and Width of the $W$
  Boson in $e^{+} e^{-}$ Collisions at $\sqrt{s}$ = 161-GeV - 209-GeV}}, Eur.
  Phys. J. C \textbf{55} (2008),
  \href{http://www.slac.stanford.edu/spires/find/hep/www?eprint=0803.2534}{1--38},
   [\href{http://arXiv.org/pdf/0803.2534}{{\tt arXiv:0803.2534}} [hep-ex]]%
\relax\mciteBstWouldAddEndPuncttrue
\mciteSetBstMidEndSepPunct{\mcitedefaultmidpunct}
{\mcitedefaultendpunct}{\mcitedefaultseppunct}\relax
\EndOfBibitem
\bibitem{ALEPH:2006cdc}
S.~Schael et~al., ALEPH, \emph{{Measurement of the $W$ boson mass and width in
  $e^{+} e^{-}$ collisions at LEP}}, Eur. Phys. J. C \textbf{47} (2006),
  \href{http://www.slac.stanford.edu/spires/find/hep/www?eprint=hep-ex/0605011}{309--335},
   [\href{http://arXiv.org/pdf/hep-ex/0605011}{{\tt hep-ex/0605011}}]%
\relax\mciteBstWouldAddEndPuncttrue
\mciteSetBstMidEndSepPunct{\mcitedefaultmidpunct}
{\mcitedefaultendpunct}{\mcitedefaultseppunct}\relax
\EndOfBibitem
\bibitem{L3:2005fft}
P.~Achard et~al., L3, \emph{{Measurement of the mass and the width of the $W$
  boson at LEP}}, Eur. Phys. J. C \textbf{45} (2006),
  \href{http://www.slac.stanford.edu/spires/find/hep/www?eprint=hep-ex/0511049}{569--587},
   [\href{http://arXiv.org/pdf/hep-ex/0511049}{{\tt hep-ex/0511049}}]%
\relax\mciteBstWouldAddEndPuncttrue
\mciteSetBstMidEndSepPunct{\mcitedefaultmidpunct}
{\mcitedefaultendpunct}{\mcitedefaultseppunct}\relax
\EndOfBibitem
\bibitem{OPAL:2005rdt}
G.~Abbiendi et~al., OPAL, \emph{{Measurement of the mass and width of the $W$
  boson}}, Eur. Phys. J. C \textbf{45} (2006),
  \href{http://www.slac.stanford.edu/spires/find/hep/www?eprint=hep-ex/0508060}{307--335},
   [\href{http://arXiv.org/pdf/hep-ex/0508060}{{\tt hep-ex/0508060}}]%
\relax\mciteBstWouldAddEndPuncttrue
\mciteSetBstMidEndSepPunct{\mcitedefaultmidpunct}
{\mcitedefaultendpunct}{\mcitedefaultseppunct}\relax
\EndOfBibitem
\bibitem{ATLAS:2016rnf}
G.~Aad et~al., ATLAS, \emph{{Measurement of the angular coefficients in
  $Z$-boson events using electron and muon pairs from data taken at
  $\sqrt{s}=8$ TeV with the ATLAS detector}}, JHEP \textbf{08} (2016),
  \href{http://www.slac.stanford.edu/spires/find/hep/www?eprint=1606.00689}{159},
   [\href{http://arXiv.org/pdf/1606.00689}{{\tt arXiv:1606.00689}} [hep-ex]]%
\relax\mciteBstWouldAddEndPuncttrue
\mciteSetBstMidEndSepPunct{\mcitedefaultmidpunct}
{\mcitedefaultendpunct}{\mcitedefaultseppunct}\relax
\EndOfBibitem
\bibitem{ATLAS:2012ewf}
G.~Aad et~al., ATLAS, \emph{{Measurement of angular correlations in Drell-Yan
  lepton pairs to probe Z/gamma* boson transverse momentum at sqrt(s)=7 TeV
  with the ATLAS detector}}, Phys. Lett. B \textbf{720} (2013),
  \href{http://www.slac.stanford.edu/spires/find/hep/www?eprint=1211.6899}{32--51},
   [\href{http://arXiv.org/pdf/1211.6899}{{\tt arXiv:1211.6899}} [hep-ex]]%
\relax\mciteBstWouldAddEndPuncttrue
\mciteSetBstMidEndSepPunct{\mcitedefaultmidpunct}
{\mcitedefaultendpunct}{\mcitedefaultseppunct}\relax
\EndOfBibitem
\bibitem{CMS:2015cyj}
V.~Khachatryan et~al., CMS, \emph{{Angular coefficients of Z bosons produced in
  pp collisions at $\sqrt{s}$ = 8 TeV and decaying to $\mu^+ \mu^-$ as a
  function of transverse momentum and rapidity}}, Phys. Lett. B \textbf{750}
  (2015),
  \href{http://www.slac.stanford.edu/spires/find/hep/www?eprint=1504.03512}{154--175},
   [\href{http://arXiv.org/pdf/1504.03512}{{\tt arXiv:1504.03512}} [hep-ex]]%
\relax\mciteBstWouldAddEndPuncttrue
\mciteSetBstMidEndSepPunct{\mcitedefaultmidpunct}
{\mcitedefaultendpunct}{\mcitedefaultseppunct}\relax
\EndOfBibitem
\bibitem{LHCb:2022tbc}
R.~Aaij et~al., LHCb, \emph{{First Measurement of the $Z\to\mu^+\mu^-$ Angular
  Coefficients in the Forward Region of $pp$ Collisions at
  $\sqrt{s}=13$\,GeV}}, Phys. Rev. Lett. \textbf{129} (2022), no.~9,
  \href{http://www.slac.stanford.edu/spires/find/hep/www?eprint=2203.01602}{091801},
   [\href{http://arXiv.org/pdf/2203.01602}{{\tt arXiv:2203.01602}} [hep-ex]]%
\relax\mciteBstWouldAddEndPuncttrue
\mciteSetBstMidEndSepPunct{\mcitedefaultmidpunct}
{\mcitedefaultendpunct}{\mcitedefaultseppunct}\relax
\EndOfBibitem
\bibitem{CDF:2011ksg}
T.~Aaltonen et~al., CDF, \emph{{First Measurement of the Angular Coefficients
  of Drell-Yan $e^{+}e^{-}$ pairs in the Z Mass Region from $p\bar{p}$
  Collisions at $\sqrt{s}$ = 1.96 TeV}}, Phys. Rev. Lett. \textbf{106} (2011),
  \href{http://www.slac.stanford.edu/spires/find/hep/www?eprint=1103.5699}{241801},
   [\href{http://arXiv.org/pdf/1103.5699}{{\tt arXiv:1103.5699}} [hep-ex]]%
\relax\mciteBstWouldAddEndPuncttrue
\mciteSetBstMidEndSepPunct{\mcitedefaultmidpunct}
{\mcitedefaultendpunct}{\mcitedefaultseppunct}\relax
\EndOfBibitem
\bibitem{CDF:2013uau}
T.~Aaltonen et~al., CDF, \emph{{Indirect Measurement of $\sin^2\theta_W$
  $(M_W)$ Using $e^+e^-$ Pairs in the Z-Boson Region with $p\bar{p}$ Collisions
  at a Center-of-Momentum Energy of 1.96 TeV}}, Phys. Rev. D \textbf{88}
  (2013), no.~7,
  \href{http://www.slac.stanford.edu/spires/find/hep/www?eprint=1307.0770}{072002},
   [\href{http://arXiv.org/pdf/1307.0770}{{\tt arXiv:1307.0770}} [hep-ex]],
  [Erratum: Phys.Rev.D 88, 079905 (2013)]%
\relax\mciteBstWouldAddEndPuncttrue
\mciteSetBstMidEndSepPunct{\mcitedefaultmidpunct}
{\mcitedefaultendpunct}{\mcitedefaultseppunct}\relax
\EndOfBibitem
\bibitem{ALEPH:2005ab}
S.~Schael et~al., ALEPH, DELPHI, L3, OPAL, SLD, LEP Electroweak Working Group,
  SLD Electroweak Group, SLD Heavy Flavour Group, \emph{{Precision electroweak
  measurements on the $Z$ resonance}}, Phys. Rept. \textbf{427} (2006),
  \href{http://www.slac.stanford.edu/spires/find/hep/www?eprint=hep-ex/0509008}{257--454},
   [\href{http://arXiv.org/pdf/hep-ex/0509008}{{\tt hep-ex/0509008}}]%
\relax\mciteBstWouldAddEndPuncttrue
\mciteSetBstMidEndSepPunct{\mcitedefaultmidpunct}
{\mcitedefaultendpunct}{\mcitedefaultseppunct}\relax
\EndOfBibitem
\bibitem{CDF:2005qwt}
D.~Acosta et~al., CDF, \emph{{Measurement of the azimuthal angle distribution
  of leptons from $W$ boson decays as a function of the $W$ transverse momentum
  in $p \bar{p}$ collisions at $\sqrt{s} = 1.8$ TeV}}, Phys. Rev. D \textbf{73}
  (2006),
  \href{http://www.slac.stanford.edu/spires/find/hep/www?eprint=hep-ex/0504020}{052002},
   [\href{http://arXiv.org/pdf/hep-ex/0504020}{{\tt hep-ex/0504020}}]%
\relax\mciteBstWouldAddEndPuncttrue
\mciteSetBstMidEndSepPunct{\mcitedefaultmidpunct}
{\mcitedefaultendpunct}{\mcitedefaultseppunct}\relax
\EndOfBibitem
\bibitem{ATLAS:2012au}
G.~Aad et~al., ATLAS, \emph{{Measurement of the polarisation of $W$ bosons
  produced with large transverse momentum in $pp$ collisions at $\sqrt{s}=7$
  TeV with the ATLAS experiment}}, Eur. Phys. J. C \textbf{72} (2012),
  \href{http://www.slac.stanford.edu/spires/find/hep/www?eprint=1203.2165}{2001},
   [\href{http://arXiv.org/pdf/1203.2165}{{\tt arXiv:1203.2165}} [hep-ex]]%
\relax\mciteBstWouldAddEndPuncttrue
\mciteSetBstMidEndSepPunct{\mcitedefaultmidpunct}
{\mcitedefaultendpunct}{\mcitedefaultseppunct}\relax
\EndOfBibitem
\bibitem{CMS:2011kaj}
S.~Chatrchyan et~al., CMS, \emph{{Measurement of the Polarization of W Bosons
  with Large Transverse Momenta in W+Jets Events at the LHC}}, Phys. Rev. Lett.
  \textbf{107} (2011),
  \href{http://www.slac.stanford.edu/spires/find/hep/www?eprint=1104.3829}{021802},
   [\href{http://arXiv.org/pdf/1104.3829}{{\tt arXiv:1104.3829}} [hep-ex]]%
\relax\mciteBstWouldAddEndPuncttrue
\mciteSetBstMidEndSepPunct{\mcitedefaultmidpunct}
{\mcitedefaultendpunct}{\mcitedefaultseppunct}\relax
\EndOfBibitem
\bibitem{Wackeroth:1996hz}
D.~Wackeroth and W.~Hollik, \emph{{Electroweak radiative corrections to
  resonant charged gauge boson production}}, Phys. Rev. D \textbf{55} (1997),
  \href{http://www.slac.stanford.edu/spires/find/hep/www?eprint=hep-ph/9606398}{6788--6818},
   [\href{http://arXiv.org/pdf/hep-ph/9606398}{{\tt hep-ph/9606398}}]%
\relax\mciteBstWouldAddEndPuncttrue
\mciteSetBstMidEndSepPunct{\mcitedefaultmidpunct}
{\mcitedefaultendpunct}{\mcitedefaultseppunct}\relax
\EndOfBibitem
\bibitem{Baur:1997wa}
U.~Baur, S.~Keller and W.~K. Sakumoto, \emph{{QED radiative corrections to $Z$
  boson production and the forward backward asymmetry at hadron colliders}},
  Phys. Rev. D \textbf{57} (1998),
  \href{http://www.slac.stanford.edu/spires/find/hep/www?eprint=hep-ph/9707301}{199--215},
   [\href{http://arXiv.org/pdf/hep-ph/9707301}{{\tt hep-ph/9707301}}]%
\relax\mciteBstWouldAddEndPuncttrue
\mciteSetBstMidEndSepPunct{\mcitedefaultmidpunct}
{\mcitedefaultendpunct}{\mcitedefaultseppunct}\relax
\EndOfBibitem
\bibitem{Baur:1998kt}
U.~Baur, S.~Keller and D.~Wackeroth, \emph{{Electroweak radiative corrections
  to $W$ boson production in hadronic collisions}}, Phys. Rev. D \textbf{59}
  (1999),
  \href{http://www.slac.stanford.edu/spires/find/hep/www?eprint=hep-ph/9807417}{013002},
   [\href{http://arXiv.org/pdf/hep-ph/9807417}{{\tt hep-ph/9807417}}]%
\relax\mciteBstWouldAddEndPuncttrue
\mciteSetBstMidEndSepPunct{\mcitedefaultmidpunct}
{\mcitedefaultendpunct}{\mcitedefaultseppunct}\relax
\EndOfBibitem
\bibitem{Baur:2001ze}
U.~Baur, O.~Brein, W.~Hollik, C.~Schappacher and D.~Wackeroth,
  \emph{{Electroweak radiative corrections to neutral current Drell-Yan
  processes at hadron colliders}}, Phys. Rev. D \textbf{65} (2002),
  \href{http://www.slac.stanford.edu/spires/find/hep/www?eprint=hep-ph/0108274}{033007},
   [\href{http://arXiv.org/pdf/hep-ph/0108274}{{\tt hep-ph/0108274}}]%
\relax\mciteBstWouldAddEndPuncttrue
\mciteSetBstMidEndSepPunct{\mcitedefaultmidpunct}
{\mcitedefaultendpunct}{\mcitedefaultseppunct}\relax
\EndOfBibitem
\bibitem{Baur:2004ig}
U.~Baur and D.~Wackeroth, \emph{{Electroweak radiative corrections to $p
  \bar{p} \to W^\pm \to \ell^\pm \nu$ beyond the pole approximation}}, Phys.
  Rev. D \textbf{70} (2004),
  \href{http://www.slac.stanford.edu/spires/find/hep/www?eprint=hep-ph/0405191}{073015},
   [\href{http://arXiv.org/pdf/hep-ph/0405191}{{\tt hep-ph/0405191}}]%
\relax\mciteBstWouldAddEndPuncttrue
\mciteSetBstMidEndSepPunct{\mcitedefaultmidpunct}
{\mcitedefaultendpunct}{\mcitedefaultseppunct}\relax
\EndOfBibitem
\bibitem{Andonov:2004hi}
A.~Andonov, A.~Arbuzov, D.~Bardin, S.~Bondarenko, P.~Christova,
  L.~Kalinovskaya, G.~Nanava and W.~von Schlippe, \emph{{SANCscope - v.1.00}},
  Comput. Phys. Commun. \textbf{174} (2006),
  \href{http://www.slac.stanford.edu/spires/find/hep/www?eprint=hep-ph/0411186}{481--517},
   [\href{http://arXiv.org/pdf/hep-ph/0411186}{{\tt hep-ph/0411186}}],
  [Erratum: Comput.Phys.Commun. 177, 623--624 (2007)]%
\relax\mciteBstWouldAddEndPuncttrue
\mciteSetBstMidEndSepPunct{\mcitedefaultmidpunct}
{\mcitedefaultendpunct}{\mcitedefaultseppunct}\relax
\EndOfBibitem
\bibitem{Dittmaier:2001ay}
S.~Dittmaier and M.~Kr\"amer, \emph{{Electroweak radiative corrections to W
  boson production at hadron colliders}}, Phys. Rev. D \textbf{65} (2002),
  \href{http://www.slac.stanford.edu/spires/find/hep/www?eprint=hep-ph/0109062}{073007},
   [\href{http://arXiv.org/pdf/hep-ph/0109062}{{\tt hep-ph/0109062}}]%
\relax\mciteBstWouldAddEndPuncttrue
\mciteSetBstMidEndSepPunct{\mcitedefaultmidpunct}
{\mcitedefaultendpunct}{\mcitedefaultseppunct}\relax
\EndOfBibitem
\bibitem{Dittmaier:2009cr}
S.~Dittmaier and M.~Huber, \emph{{Radiative corrections to the neutral-current
  Drell-Yan process in the Standard Model and its minimal supersymmetric
  extension}}, JHEP \textbf{01} (2010),
  \href{http://www.slac.stanford.edu/spires/find/hep/www?eprint=0911.2329}{060},
   [\href{http://arXiv.org/pdf/0911.2329}{{\tt arXiv:0911.2329}} [hep-ph]]%
\relax\mciteBstWouldAddEndPuncttrue
\mciteSetBstMidEndSepPunct{\mcitedefaultmidpunct}
{\mcitedefaultendpunct}{\mcitedefaultseppunct}\relax
\EndOfBibitem
\bibitem{Li:2012wna}
Y.~Li and F.~Petriello, \emph{{Combining QCD and electroweak corrections to
  dilepton production in FEWZ}}, Phys. Rev. D \textbf{86} (2012),
  \href{http://www.slac.stanford.edu/spires/find/hep/www?eprint=1208.5967}{094034},
   [\href{http://arXiv.org/pdf/1208.5967}{{\tt arXiv:1208.5967}} [hep-ph]]%
\relax\mciteBstWouldAddEndPuncttrue
\mciteSetBstMidEndSepPunct{\mcitedefaultmidpunct}
{\mcitedefaultendpunct}{\mcitedefaultseppunct}\relax
\EndOfBibitem
\bibitem{Alioli:2016fum}
S.~Alioli et~al., \emph{{Precision studies of observables in $p p \rightarrow W
  \rightarrow l\nu _l$ and $ pp \rightarrow \gamma ,Z \rightarrow l^+ l^-$
  processes at the LHC}}, Eur. Phys. J. \textbf{C77} (2017), no.~5,
  \href{http://www.slac.stanford.edu/spires/find/hep/www?eprint=1606.02330}{280},
   [\href{http://arXiv.org/pdf/1606.02330}{{\tt arXiv:1606.02330}} [hep-ph]]%
\relax\mciteBstWouldAddEndPuncttrue
\mciteSetBstMidEndSepPunct{\mcitedefaultmidpunct}
{\mcitedefaultendpunct}{\mcitedefaultseppunct}\relax
\EndOfBibitem
\bibitem{Bonciani:2021zzf}
R.~Bonciani, L.~Buonocore, M.~Grazzini, S.~Kallweit, N.~Rana, F.~Tramontano and
  A.~Vicini, \emph{{Mixed Strong-Electroweak Corrections to the Drell-Yan
  Process}}, Phys. Rev. Lett. \textbf{128} (2022), no.~1,
  \href{http://www.slac.stanford.edu/spires/find/hep/www?eprint=2106.11953}{012002},
   [\href{http://arXiv.org/pdf/2106.11953}{{\tt arXiv:2106.11953}} [hep-ph]]%
\relax\mciteBstWouldAddEndPuncttrue
\mciteSetBstMidEndSepPunct{\mcitedefaultmidpunct}
{\mcitedefaultendpunct}{\mcitedefaultseppunct}\relax
\EndOfBibitem
\bibitem{Buonocore:2021rxx}
L.~Buonocore, M.~Grazzini, S.~Kallweit, C.~Savoini and F.~Tramontano,
  \emph{{Mixed QCD-EW corrections to $\boldsymbol{pp\!\to\!\ell\nu_\ell\!+\!X}$
  at the LHC}}, Phys. Rev. D \textbf{103} (2021),
  \href{http://www.slac.stanford.edu/spires/find/hep/www?eprint=2102.12539}{114012},
   [\href{http://arXiv.org/pdf/2102.12539}{{\tt arXiv:2102.12539}} [hep-ph]]%
\relax\mciteBstWouldAddEndPuncttrue
\mciteSetBstMidEndSepPunct{\mcitedefaultmidpunct}
{\mcitedefaultendpunct}{\mcitedefaultseppunct}\relax
\EndOfBibitem
\bibitem{Armadillo:2022bgm}
T.~Armadillo, R.~Bonciani, S.~Devoto, N.~Rana and A.~Vicini, \emph{{Two-loop
  mixed QCD-EW corrections to neutral current Drell-Yan}}, JHEP \textbf{05}
  (2022),
  \href{http://www.slac.stanford.edu/spires/find/hep/www?eprint=2201.01754}{072},
   [\href{http://arXiv.org/pdf/2201.01754}{{\tt arXiv:2201.01754}} [hep-ph]]%
\relax\mciteBstWouldAddEndPuncttrue
\mciteSetBstMidEndSepPunct{\mcitedefaultmidpunct}
{\mcitedefaultendpunct}{\mcitedefaultseppunct}\relax
\EndOfBibitem
\bibitem{Delto:2019ewv}
M.~Delto, M.~Jaquier, K.~Melnikov and R.~R\"ontsch, \emph{{Mixed
  QCD$\otimes$QED corrections to on-shell $Z$ boson production at the LHC}},
  JHEP \textbf{01} (2020),
  \href{http://www.slac.stanford.edu/spires/find/hep/www?eprint=1909.08428}{043},
   [\href{http://arXiv.org/pdf/1909.08428}{{\tt arXiv:1909.08428}} [hep-ph]]%
\relax\mciteBstWouldAddEndPuncttrue
\mciteSetBstMidEndSepPunct{\mcitedefaultmidpunct}
{\mcitedefaultendpunct}{\mcitedefaultseppunct}\relax
\EndOfBibitem
\bibitem{Buccioni:2020cfi}
F.~Buccioni, F.~Caola, M.~Delto, M.~Jaquier, K.~Melnikov and R.~R\"ontsch,
  \emph{{Mixed QCD-electroweak corrections to on-shell Z production at the
  LHC}}, Phys. Lett. B \textbf{811} (2020),
  \href{http://www.slac.stanford.edu/spires/find/hep/www?eprint=2005.10221}{135969},
   [\href{http://arXiv.org/pdf/2005.10221}{{\tt arXiv:2005.10221}} [hep-ph]]%
\relax\mciteBstWouldAddEndPuncttrue
\mciteSetBstMidEndSepPunct{\mcitedefaultmidpunct}
{\mcitedefaultendpunct}{\mcitedefaultseppunct}\relax
\EndOfBibitem
\bibitem{Bonciani:2020tvf}
R.~Bonciani, F.~Buccioni, N.~Rana and A.~Vicini, \emph{{Next-to-Next-to-Leading
  Order Mixed QCD-Electroweak Corrections to on-Shell Z Production}}, Phys.
  Rev. Lett. \textbf{125} (2020), no.~23,
  \href{http://www.slac.stanford.edu/spires/find/hep/www?eprint=2007.06518}{232004},
   [\href{http://arXiv.org/pdf/2007.06518}{{\tt arXiv:2007.06518}} [hep-ph]]%
\relax\mciteBstWouldAddEndPuncttrue
\mciteSetBstMidEndSepPunct{\mcitedefaultmidpunct}
{\mcitedefaultendpunct}{\mcitedefaultseppunct}\relax
\EndOfBibitem
\bibitem{Bonciani:2021iis}
R.~Bonciani, F.~Buccioni, N.~Rana and A.~Vicini, \emph{{On-shell Z boson
  production at hadron colliders through $\mathcal{O}(\alpha\alpha_{s})$}},
  JHEP \textbf{02} (2022),
  \href{http://www.slac.stanford.edu/spires/find/hep/www?eprint=2111.12694}{095},
   [\href{http://arXiv.org/pdf/2111.12694}{{\tt arXiv:2111.12694}} [hep-ph]]%
\relax\mciteBstWouldAddEndPuncttrue
\mciteSetBstMidEndSepPunct{\mcitedefaultmidpunct}
{\mcitedefaultendpunct}{\mcitedefaultseppunct}\relax
\EndOfBibitem
\bibitem{Behring:2020cqi}
A.~Behring, F.~Buccioni, F.~Caola, M.~Delto, M.~Jaquier, K.~Melnikov and
  R.~R\"ontsch, \emph{{Mixed QCD-electroweak corrections to $W$-boson
  production in hadron collisions}}, Phys. Rev. D \textbf{103} (2021), no.~1,
  \href{http://www.slac.stanford.edu/spires/find/hep/www?eprint=2009.10386}{013008},
   [\href{http://arXiv.org/pdf/2009.10386}{{\tt arXiv:2009.10386}} [hep-ph]]%
\relax\mciteBstWouldAddEndPuncttrue
\mciteSetBstMidEndSepPunct{\mcitedefaultmidpunct}
{\mcitedefaultendpunct}{\mcitedefaultseppunct}\relax
\EndOfBibitem
\bibitem{Dittmaier:2014qza}
S.~Dittmaier, A.~Huss and C.~Schwinn, \emph{{Mixed QCD-electroweak
  $\mathcal{O}(\alpha_s\alpha)$ corrections to Drell-Yan processes in the
  resonance region: pole approximation and non-factorizable corrections}},
  Nucl. Phys. B \textbf{885} (2014),
  \href{http://www.slac.stanford.edu/spires/find/hep/www?eprint=1403.3216}{318--372},
   [\href{http://arXiv.org/pdf/1403.3216}{{\tt arXiv:1403.3216}} [hep-ph]]%
\relax\mciteBstWouldAddEndPuncttrue
\mciteSetBstMidEndSepPunct{\mcitedefaultmidpunct}
{\mcitedefaultendpunct}{\mcitedefaultseppunct}\relax
\EndOfBibitem
\bibitem{Dittmaier:2015rxo}
S.~Dittmaier, A.~Huss and C.~Schwinn, \emph{{Dominant mixed QCD-electroweak
  O($\alpha_s\alpha$) corrections to Drell\textendash{}Yan processes in the
  resonance region}}, Nucl. Phys. B \textbf{904} (2016),
  \href{http://www.slac.stanford.edu/spires/find/hep/www?eprint=1511.08016}{216--252},
   [\href{http://arXiv.org/pdf/1511.08016}{{\tt arXiv:1511.08016}} [hep-ph]]%
\relax\mciteBstWouldAddEndPuncttrue
\mciteSetBstMidEndSepPunct{\mcitedefaultmidpunct}
{\mcitedefaultendpunct}{\mcitedefaultseppunct}\relax
\EndOfBibitem
\bibitem{Seymour:1991xa}
M.~H. Seymour, \emph{{Photon radiation in final state parton showering}}, Z.
  Phys. C \textbf{56} (1992),
  \href{http://www.slac.stanford.edu/spires/find/hep/www?j=Z%20Phys%20C,56,161}{161--170},
  CAVENDISH-HEP-91-16%
\relax\mciteBstWouldAddEndPuncttrue
\mciteSetBstMidEndSepPunct{\mcitedefaultmidpunct}
{\mcitedefaultendpunct}{\mcitedefaultseppunct}\relax
\EndOfBibitem
\bibitem{Yennie:1961ad}
D.~R. Yennie, S.~C. Frautschi and H.~Suura, \emph{{The infrared divergence
  phenomena and high-energy processes}}, Annals Phys. \textbf{13} (1961),
  \href{http://www.slac.stanford.edu/spires/find/hep/www?j=Annals%20Phys,13,379}{379--452}%
\relax\mciteBstWouldAddEndPuncttrue
\mciteSetBstMidEndSepPunct{\mcitedefaultmidpunct}
{\mcitedefaultendpunct}{\mcitedefaultseppunct}\relax
\EndOfBibitem
\bibitem{Bellm:2019zci}
J.~Bellm et~al., \emph{{Herwig 7.2 release note}}, Eur. Phys. J. C \textbf{80}
  (2020), no.~5,
  \href{http://www.slac.stanford.edu/spires/find/hep/www?eprint=1912.06509}{452},
   [\href{http://arXiv.org/pdf/1912.06509}{{\tt arXiv:1912.06509}} [hep-ph]]%
\relax\mciteBstWouldAddEndPuncttrue
\mciteSetBstMidEndSepPunct{\mcitedefaultmidpunct}
{\mcitedefaultendpunct}{\mcitedefaultseppunct}\relax
\EndOfBibitem
\bibitem{Bellm:2015jjp}
J.~Bellm et~al., \emph{{Herwig 7.0/Herwig++ 3.0 release note}}, Eur. Phys. J. C
  \textbf{76} (2016), no.~4,
  \href{http://www.slac.stanford.edu/spires/find/hep/www?eprint=1512.01178}{196},
   [\href{http://arXiv.org/pdf/1512.01178}{{\tt arXiv:1512.01178}} [hep-ph]]%
\relax\mciteBstWouldAddEndPuncttrue
\mciteSetBstMidEndSepPunct{\mcitedefaultmidpunct}
{\mcitedefaultendpunct}{\mcitedefaultseppunct}\relax
\EndOfBibitem
\bibitem{Bierlich:2022pfr}
\href{http://www.slac.stanford.edu/spires/find/hep/www?eprint=2203.11601}{C.~Bierlich
  et~al.}, \emph{{A comprehensive guide to the physics and usage of PYTHIA
  8.3}},  \href{http://arXiv.org/pdf/2203.11601}{{\tt arXiv:2203.11601}}
  [hep-ph]%
\relax\mciteBstWouldAddEndPuncttrue
\mciteSetBstMidEndSepPunct{\mcitedefaultmidpunct}
{\mcitedefaultendpunct}{\mcitedefaultseppunct}\relax
\EndOfBibitem
\bibitem{Sjostrand:2014zea}
T.~Sj{\"o}strand, S.~Ask, J.~R. Christiansen, R.~Corke, N.~Desai, P.~Ilten,
  S.~Mrenna, S.~Prestel, C.~O. Rasmussen and P.~Z. Skands, \emph{{An
  introduction to PYTHIA 8.2}}, Comput. Phys. Commun. \textbf{191} (2015),
  \href{http://www.slac.stanford.edu/spires/find/hep/www?eprint=1410.3012}{159--177},
   [\href{http://arXiv.org/pdf/1410.3012}{{\tt arXiv:1410.3012}} [hep-ph]]%
\relax\mciteBstWouldAddEndPuncttrue
\mciteSetBstMidEndSepPunct{\mcitedefaultmidpunct}
{\mcitedefaultendpunct}{\mcitedefaultseppunct}\relax
\EndOfBibitem
\bibitem{Hoeche:2009xc}
S.~H{\"o}che, S.~Schumann and F.~Siegert, \emph{{Hard photon production and
  matrix-element parton-shower merging}}, Phys. Rev. D \textbf{81} (2010),
  \href{http://www.slac.stanford.edu/spires/find/hep/www?eprint=0912.3501}{034026},
   [\href{http://arXiv.org/pdf/0912.3501}{{\tt arXiv:0912.3501}} [hep-ph]]%
\relax\mciteBstWouldAddEndPuncttrue
\mciteSetBstMidEndSepPunct{\mcitedefaultmidpunct}
{\mcitedefaultendpunct}{\mcitedefaultseppunct}\relax
\EndOfBibitem
\bibitem{Gleisberg:2008ta}
T.~Gleisberg, S.~H{\"o}che, F.~Krauss, M.~Sch{\"o}nherr, S.~Schumann,
  F.~Siegert and J.~Winter, \emph{{Event generation with SHERPA 1.1}}, JHEP
  \textbf{02} (2009),
  \href{http://www.slac.stanford.edu/spires/find/hep/www?eprint=0811.4622}{007},
   [\href{http://arXiv.org/pdf/0811.4622}{{\tt arXiv:0811.4622}} [hep-ph]]%
\relax\mciteBstWouldAddEndPuncttrue
\mciteSetBstMidEndSepPunct{\mcitedefaultmidpunct}
{\mcitedefaultendpunct}{\mcitedefaultseppunct}\relax
\EndOfBibitem
\bibitem{Bothmann:2019yzt}
E.~Bothmann et~al., \emph{{Event Generation with SHERPA 2.2}}, 2019%
\relax\mciteBstWouldAddEndPuncttrue
\mciteSetBstMidEndSepPunct{\mcitedefaultmidpunct}
{\mcitedefaultendpunct}{\mcitedefaultseppunct}\relax
\EndOfBibitem
\bibitem{Hamilton:2006xz}
K.~Hamilton and P.~Richardson, \emph{{Simulation of QED radiation in particle
  decays using the YFS formalism}}, JHEP \textbf{07} (2006),
  \href{http://www.slac.stanford.edu/spires/find/hep/www?eprint=hep-ph/0603034}{010},
   [\href{http://arXiv.org/pdf/hep-ph/0603034}{{\tt hep-ph/0603034}}]%
\relax\mciteBstWouldAddEndPuncttrue
\mciteSetBstMidEndSepPunct{\mcitedefaultmidpunct}
{\mcitedefaultendpunct}{\mcitedefaultseppunct}\relax
\EndOfBibitem
\bibitem{Schonherr:2008av}
M.~Sch{\"o}nherr and F.~Krauss, \emph{{Soft Photon Radiation in Particle Decays
  in SHERPA}}, JHEP \textbf{12} (2008),
  \href{http://www.slac.stanford.edu/spires/find/hep/www?eprint=0810.5071}{018},
   [\href{http://arXiv.org/pdf/0810.5071}{{\tt arXiv:0810.5071}} [hep-ph]]%
\relax\mciteBstWouldAddEndPuncttrue
\mciteSetBstMidEndSepPunct{\mcitedefaultmidpunct}
{\mcitedefaultendpunct}{\mcitedefaultseppunct}\relax
\EndOfBibitem
\bibitem{Krauss:2018djz}
F.~Krauss, J.~M. Lindert, R.~Linten and M.~Sch{\"o}nherr, \emph{{Accurate
  simulation of W, Z and Higgs boson decays in Sherpa}}, Eur. Phys. J. C
  \textbf{79} (2019), no.~2,
  \href{http://www.slac.stanford.edu/spires/find/hep/www?eprint=1809.10650}{143},
   [\href{http://arXiv.org/pdf/1809.10650}{{\tt arXiv:1809.10650}} [hep-ph]]%
\relax\mciteBstWouldAddEndPuncttrue
\mciteSetBstMidEndSepPunct{\mcitedefaultmidpunct}
{\mcitedefaultendpunct}{\mcitedefaultseppunct}\relax
\EndOfBibitem
\bibitem{Krauss:2022ajk}
F.~Krauss, A.~Price and M.~Sch\"onherr, \emph{{YFS Resummation for Future
  Lepton-Lepton Colliders in SHERPA}}, SciPost Phys. \textbf{13} (2022), no.~2,
  \href{http://www.slac.stanford.edu/spires/find/hep/www?eprint=2203.10948}{026},
   [\href{http://arXiv.org/pdf/2203.10948}{{\tt arXiv:2203.10948}} [hep-ph]]%
\relax\mciteBstWouldAddEndPuncttrue
\mciteSetBstMidEndSepPunct{\mcitedefaultmidpunct}
{\mcitedefaultendpunct}{\mcitedefaultseppunct}\relax
\EndOfBibitem
\bibitem{CarloniCalame:2001ny}
C.~M. Carloni~Calame, \emph{{An Improved parton shower algorithm in QED}},
  Phys. Lett. B \textbf{520} (2001),
  \href{http://www.slac.stanford.edu/spires/find/hep/www?eprint=hep-ph/0103117}{16--24},
   [\href{http://arXiv.org/pdf/hep-ph/0103117}{{\tt hep-ph/0103117}}]%
\relax\mciteBstWouldAddEndPuncttrue
\mciteSetBstMidEndSepPunct{\mcitedefaultmidpunct}
{\mcitedefaultendpunct}{\mcitedefaultseppunct}\relax
\EndOfBibitem
\bibitem{CarloniCalame:2003ux}
C.~Carloni~Calame, G.~Montagna, O.~Nicrosini and M.~Treccani, \emph{{Higher
  order QED corrections to W boson mass determination at hadron colliders}},
  Phys. Rev. D \textbf{69} (2004),
  \href{http://www.slac.stanford.edu/spires/find/hep/www?eprint=hep-ph/0303102}{037301},
   [\href{http://arXiv.org/pdf/hep-ph/0303102}{{\tt hep-ph/0303102}}]%
\relax\mciteBstWouldAddEndPuncttrue
\mciteSetBstMidEndSepPunct{\mcitedefaultmidpunct}
{\mcitedefaultendpunct}{\mcitedefaultseppunct}\relax
\EndOfBibitem
\bibitem{CarloniCalame:2005vc}
C.~M. Carloni~Calame, G.~Montagna, O.~Nicrosini and M.~Treccani,
  \emph{{Multiple photon corrections to the neutral-current Drell-Yan
  process}}, JHEP \textbf{05} (2005),
  \href{http://www.slac.stanford.edu/spires/find/hep/www?eprint=hep-ph/0502218}{019},
   [\href{http://arXiv.org/pdf/hep-ph/0502218}{{\tt hep-ph/0502218}}]%
\relax\mciteBstWouldAddEndPuncttrue
\mciteSetBstMidEndSepPunct{\mcitedefaultmidpunct}
{\mcitedefaultendpunct}{\mcitedefaultseppunct}\relax
\EndOfBibitem
\bibitem{CarloniCalame:2006zq}
C.~Carloni~Calame, G.~Montagna, O.~Nicrosini and A.~Vicini, \emph{{Precision
  electroweak calculation of the charged current Drell-Yan process}}, JHEP
  \textbf{12} (2006),
  \href{http://www.slac.stanford.edu/spires/find/hep/www?eprint=hep-ph/0609170}{016},
   [\href{http://arXiv.org/pdf/hep-ph/0609170}{{\tt hep-ph/0609170}}]%
\relax\mciteBstWouldAddEndPuncttrue
\mciteSetBstMidEndSepPunct{\mcitedefaultmidpunct}
{\mcitedefaultendpunct}{\mcitedefaultseppunct}\relax
\EndOfBibitem
\bibitem{CarloniCalame:2007cd}
C.~M. Carloni~Calame, G.~Montagna, O.~Nicrosini and A.~Vicini, \emph{{Precision
  electroweak calculation of the production of a high transverse-momentum
  lepton pair at hadron colliders}}, JHEP \textbf{10} (2007),
  \href{http://www.slac.stanford.edu/spires/find/hep/www?eprint=0710.1722}{109},
   [\href{http://arXiv.org/pdf/0710.1722}{{\tt arXiv:0710.1722}} [hep-ph]]%
\relax\mciteBstWouldAddEndPuncttrue
\mciteSetBstMidEndSepPunct{\mcitedefaultmidpunct}
{\mcitedefaultendpunct}{\mcitedefaultseppunct}\relax
\EndOfBibitem
\bibitem{CarloniCalame:2016ouw}
C.~M. Carloni~Calame, M.~Chiesa, H.~Martinez, G.~Montagna, O.~Nicrosini,
  F.~Piccinini and A.~Vicini, \emph{{Precision Measurement of the W-Boson Mass:
  Theoretical Contributions and Uncertainties}}, Phys. Rev. D \textbf{96}
  (2017), no.~9,
  \href{http://www.slac.stanford.edu/spires/find/hep/www?eprint=1612.02841}{093005},
   [\href{http://arXiv.org/pdf/1612.02841}{{\tt arXiv:1612.02841}} [hep-ph]]%
\relax\mciteBstWouldAddEndPuncttrue
\mciteSetBstMidEndSepPunct{\mcitedefaultmidpunct}
{\mcitedefaultendpunct}{\mcitedefaultseppunct}\relax
\EndOfBibitem
\bibitem{Bernaciak:2012hj}
C.~Bernaciak and D.~Wackeroth, \emph{{Combining NLO QCD and Electroweak
  Radiative Corrections to W boson Production at Hadron Colliders in the POWHEG
  Framework}}, Phys. Rev. D \textbf{85} (2012),
  \href{http://www.slac.stanford.edu/spires/find/hep/www?eprint=1201.4804}{093003},
   [\href{http://arXiv.org/pdf/1201.4804}{{\tt arXiv:1201.4804}} [hep-ph]]%
\relax\mciteBstWouldAddEndPuncttrue
\mciteSetBstMidEndSepPunct{\mcitedefaultmidpunct}
{\mcitedefaultendpunct}{\mcitedefaultseppunct}\relax
\EndOfBibitem
\bibitem{Barze:2012tt}
L.~Barze, G.~Montagna, P.~Nason, O.~Nicrosini and F.~Piccinini,
  \emph{{Implementation of electroweak corrections in the POWHEG BOX: single W
  production}}, JHEP \textbf{04} (2012),
  \href{http://www.slac.stanford.edu/spires/find/hep/www?eprint=1202.0465}{037},
   [\href{http://arXiv.org/pdf/1202.0465}{{\tt arXiv:1202.0465}} [hep-ph]]%
\relax\mciteBstWouldAddEndPuncttrue
\mciteSetBstMidEndSepPunct{\mcitedefaultmidpunct}
{\mcitedefaultendpunct}{\mcitedefaultseppunct}\relax
\EndOfBibitem
\bibitem{Muck:2016pko}
A.~M\"uck and L.~Oymanns, \emph{{Resonace-improved parton-shower matching for
  the Drell-Yan process including electroweak corrections}}, JHEP \textbf{05}
  (2017),
  \href{http://www.slac.stanford.edu/spires/find/hep/www?eprint=1612.04292}{090},
   [\href{http://arXiv.org/pdf/1612.04292}{{\tt arXiv:1612.04292}} [hep-ph]]%
\relax\mciteBstWouldAddEndPuncttrue
\mciteSetBstMidEndSepPunct{\mcitedefaultmidpunct}
{\mcitedefaultendpunct}{\mcitedefaultseppunct}\relax
\EndOfBibitem
\bibitem{Barze:2013fru}
L.~Barze, G.~Montagna, P.~Nason, O.~Nicrosini, F.~Piccinini and A.~Vicini,
  \emph{{Neutral current Drell-Yan with combined QCD and electroweak
  corrections in the POWHEG BOX}}, Eur. Phys. J. C \textbf{73} (2013), no.~6,
  \href{http://www.slac.stanford.edu/spires/find/hep/www?eprint=1302.4606}{2474},
   [\href{http://arXiv.org/pdf/1302.4606}{{\tt arXiv:1302.4606}} [hep-ph]]%
\relax\mciteBstWouldAddEndPuncttrue
\mciteSetBstMidEndSepPunct{\mcitedefaultmidpunct}
{\mcitedefaultendpunct}{\mcitedefaultseppunct}\relax
\EndOfBibitem
\bibitem{Placzek:2003zg}
W.~Placzek and S.~Jadach, \emph{{Multiphoton radiation in leptonic W boson
  decays}}, Eur. Phys. J. C \textbf{29} (2003),
  \href{http://www.slac.stanford.edu/spires/find/hep/www?eprint=hep-ph/0302065}{325--339},
   [\href{http://arXiv.org/pdf/hep-ph/0302065}{{\tt hep-ph/0302065}}]%
\relax\mciteBstWouldAddEndPuncttrue
\mciteSetBstMidEndSepPunct{\mcitedefaultmidpunct}
{\mcitedefaultendpunct}{\mcitedefaultseppunct}\relax
\EndOfBibitem
\bibitem{Barberio:1990ms}
E.~Barberio, B.~van Eijk and Z.~Was, \emph{{PHOTOS: A Universal Monte Carlo for
  QED radiative corrections in decays}}, Comput. Phys. Commun. \textbf{66}
  (1991),
  \href{http://www.slac.stanford.edu/spires/find/hep/www?j=Comput%20Phys%20Commun,66,115}{115--128},
  CERN-TH-5857-90%
\relax\mciteBstWouldAddEndPuncttrue
\mciteSetBstMidEndSepPunct{\mcitedefaultmidpunct}
{\mcitedefaultendpunct}{\mcitedefaultseppunct}\relax
\EndOfBibitem
\bibitem{Barberio:1993qi}
E.~Barberio and Z.~Was, \emph{{PHOTOS: A Universal Monte Carlo for QED
  radiative corrections. Version 2.0}}, Comput. Phys. Commun. \textbf{79}
  (1994),
  \href{http://www.slac.stanford.edu/spires/find/hep/www?j=Comput%20Phys%20Commun,79,291}{291--308},
  CERN-TH-7033-93%
\relax\mciteBstWouldAddEndPuncttrue
\mciteSetBstMidEndSepPunct{\mcitedefaultmidpunct}
{\mcitedefaultendpunct}{\mcitedefaultseppunct}\relax
\EndOfBibitem
\bibitem{Golonka:2005pn}
P.~Golonka and Z.~Was, \emph{{PHOTOS Monte Carlo: A Precision tool for QED
  corrections in $Z$ and $W$ decays}}, Eur. Phys. J. C \textbf{45} (2006),
  \href{http://www.slac.stanford.edu/spires/find/hep/www?eprint=hep-ph/0506026}{97--107},
   [\href{http://arXiv.org/pdf/hep-ph/0506026}{{\tt hep-ph/0506026}}]%
\relax\mciteBstWouldAddEndPuncttrue
\mciteSetBstMidEndSepPunct{\mcitedefaultmidpunct}
{\mcitedefaultendpunct}{\mcitedefaultseppunct}\relax
\EndOfBibitem
\bibitem{Davidson:2010ew}
N.~Davidson, T.~Przedzinski and Z.~Was, \emph{{PHOTOS interface in C++:
  Technical and Physics Documentation}}, Comput. Phys. Commun. \textbf{199}
  (2016),
  \href{http://www.slac.stanford.edu/spires/find/hep/www?eprint=1011.0937}{86--101},
   [\href{http://arXiv.org/pdf/1011.0937}{{\tt arXiv:1011.0937}} [hep-ph]]%
\relax\mciteBstWouldAddEndPuncttrue
\mciteSetBstMidEndSepPunct{\mcitedefaultmidpunct}
{\mcitedefaultendpunct}{\mcitedefaultseppunct}\relax
\EndOfBibitem
\bibitem{Gutschow:2020cug}
C.~G\"utschow and M.~Sch\"onherr, \emph{{Four lepton production and the
  accuracy of QED FSR}}, Eur. Phys. J. C \textbf{81} (2021), no.~1,
  \href{http://www.slac.stanford.edu/spires/find/hep/www?eprint=2007.15360}{48},
   [\href{http://arXiv.org/pdf/2007.15360}{{\tt arXiv:2007.15360}} [hep-ph]]%
\relax\mciteBstWouldAddEndPuncttrue
\mciteSetBstMidEndSepPunct{\mcitedefaultmidpunct}
{\mcitedefaultendpunct}{\mcitedefaultseppunct}\relax
\EndOfBibitem
\bibitem{Kotwal:2015bfa}
A.~V. Kotwal and B.~Jayatilaka, \emph{{Comparison of HORACE and PHOTOS
  Algorithms for Multiphoton Emission in the Context of $W$ Boson Mass
  Measurement}}, Adv. High Energy Phys. \textbf{2016} (2016),
  \href{http://www.slac.stanford.edu/spires/find/hep/www?eprint=1510.02458}{1615081},
   [\href{http://arXiv.org/pdf/1510.02458}{{\tt arXiv:1510.02458}} [hep-ph]]%
\relax\mciteBstWouldAddEndPuncttrue
\mciteSetBstMidEndSepPunct{\mcitedefaultmidpunct}
{\mcitedefaultendpunct}{\mcitedefaultseppunct}\relax
\EndOfBibitem
\bibitem{CarloniCalame:2004fza}
C.~M. Carloni~Calame, S.~Jadach, G.~Montagna, O.~Nicrosini and W.~Placzek,
  \emph{{Comparisons of the Monte Carlo programs HORACE and WINHAC for single W
  boson production at hadron colliders}}, Acta Phys. Polon. B \textbf{35}
  (2004),
  \href{http://www.slac.stanford.edu/spires/find/hep/www?eprint=hep-ph/0402235}{1643--1674},
   [\href{http://arXiv.org/pdf/hep-ph/0402235}{{\tt hep-ph/0402235}}]%
\relax\mciteBstWouldAddEndPuncttrue
\mciteSetBstMidEndSepPunct{\mcitedefaultmidpunct}
{\mcitedefaultendpunct}{\mcitedefaultseppunct}\relax
\EndOfBibitem
\bibitem{Arbuzov:2012dx}
A.~Arbuzov, R.~Sadykov and Z.~Was, \emph{{QED Bremsstrahlung in decays of
  electroweak bosons}}, Eur. Phys. J. C \textbf{73} (2013), no.~11,
  \href{http://www.slac.stanford.edu/spires/find/hep/www?eprint=1212.6783}{2625},
   [\href{http://arXiv.org/pdf/1212.6783}{{\tt arXiv:1212.6783}} [hep-ph]]%
\relax\mciteBstWouldAddEndPuncttrue
\mciteSetBstMidEndSepPunct{\mcitedefaultmidpunct}
{\mcitedefaultendpunct}{\mcitedefaultseppunct}\relax
\EndOfBibitem
\bibitem{Antropov:2017bed}
S.~Antropov, A.~Arbuzov, R.~Sadykov and Z.~Was, \emph{{Extra lepton pair
  emission corrections to Drell-Yan processes in PHOTOS and SANC}}, Acta Phys.
  Polon. B \textbf{48} (2017),
  \href{http://www.slac.stanford.edu/spires/find/hep/www?eprint=1706.05571}{1469},
   [\href{http://arXiv.org/pdf/1706.05571}{{\tt arXiv:1706.05571}} [hep-ph]]%
\relax\mciteBstWouldAddEndPuncttrue
\mciteSetBstMidEndSepPunct{\mcitedefaultmidpunct}
{\mcitedefaultendpunct}{\mcitedefaultseppunct}\relax
\EndOfBibitem
\bibitem{Schumann:2007mg}
S.~Schumann and F.~Krauss, \emph{{A Parton shower algorithm based on
  Catani-Seymour dipole factorisation}}, JHEP \textbf{03} (2008),
  \href{http://www.slac.stanford.edu/spires/find/hep/www?eprint=0709.1027}{038},
   [\href{http://arXiv.org/pdf/0709.1027}{{\tt arXiv:0709.1027}} [hep-ph]]%
\relax\mciteBstWouldAddEndPuncttrue
\mciteSetBstMidEndSepPunct{\mcitedefaultmidpunct}
{\mcitedefaultendpunct}{\mcitedefaultseppunct}\relax
\EndOfBibitem
\bibitem{Seymour:1994df}
M.~H. Seymour, \emph{{Matrix element corrections to parton shower algorithms}},
  Comput. Phys. Commun. \textbf{90} (1995),
  \href{http://www.slac.stanford.edu/spires/find/hep/www?eprint=hep-ph/9410414}{95--101},
   [\href{http://arXiv.org/pdf/hep-ph/9410414}{{\tt hep-ph/9410414}}]%
\relax\mciteBstWouldAddEndPuncttrue
\mciteSetBstMidEndSepPunct{\mcitedefaultmidpunct}
{\mcitedefaultendpunct}{\mcitedefaultseppunct}\relax
\EndOfBibitem
\bibitem{Sjostrand:2006za}
T.~Sjostrand, S.~Mrenna and P.~Z. Skands, \emph{{PYTHIA 6.4 Physics and
  Manual}}, JHEP \textbf{05} (2006),
  \href{http://www.slac.stanford.edu/spires/find/hep/www?eprint=hep-ph/0603175}{026},
   [\href{http://arXiv.org/pdf/hep-ph/0603175}{{\tt hep-ph/0603175}}]%
\relax\mciteBstWouldAddEndPuncttrue
\mciteSetBstMidEndSepPunct{\mcitedefaultmidpunct}
{\mcitedefaultendpunct}{\mcitedefaultseppunct}\relax
\EndOfBibitem
\bibitem{Catani:2002hc}
S.~Catani, S.~Dittmaier, M.~H. Seymour and Z.~Trocsanyi, \emph{{The Dipole
  formalism for next-to-leading order QCD calculations with massive partons}},
  Nucl. Phys. \textbf{B627} (2002),
  \href{http://www.slac.stanford.edu/spires/find/hep/www?eprint=hep-ph/0201036}{189--265},
   [\href{http://arXiv.org/pdf/hep-ph/0201036}{{\tt arXiv:hep-ph/0201036}}
  [hep-ph]]%
\relax\mciteBstWouldAddEndPuncttrue
\mciteSetBstMidEndSepPunct{\mcitedefaultmidpunct}
{\mcitedefaultendpunct}{\mcitedefaultseppunct}\relax
\EndOfBibitem
\bibitem{Hoche:2014kca}
S.~H\"oche, S.~Kuttimalai, S.~Schumann and F.~Siegert, \emph{{Beyond Standard
  Model calculations with Sherpa}}, Eur. Phys. J. C \textbf{75} (2015), no.~3,
  \href{http://www.slac.stanford.edu/spires/find/hep/www?eprint=1412.6478}{135},
   [\href{http://arXiv.org/pdf/1412.6478}{{\tt arXiv:1412.6478}} [hep-ph]]%
\relax\mciteBstWouldAddEndPuncttrue
\mciteSetBstMidEndSepPunct{\mcitedefaultmidpunct}
{\mcitedefaultendpunct}{\mcitedefaultseppunct}\relax
\EndOfBibitem
\bibitem{Schonherr:2017qcj}
M.~Sch{\"o}nherr, \emph{{An automated subtraction of NLO EW infrared
  divergences}}, Eur. Phys. J. \textbf{C78} (2018), no.~2,
  \href{http://www.slac.stanford.edu/spires/find/hep/www?eprint=1712.07975}{119},
   [\href{http://arXiv.org/pdf/1712.07975}{{\tt arXiv:1712.07975}} [hep-ph]]%
\relax\mciteBstWouldAddEndPuncttrue
\mciteSetBstMidEndSepPunct{\mcitedefaultmidpunct}
{\mcitedefaultendpunct}{\mcitedefaultseppunct}\relax
\EndOfBibitem
\bibitem{Dittmaier:2008md}
S.~Dittmaier, A.~Kabelschacht and T.~Kasprzik, \emph{{Polarized QED splittings
  of massive fermions and dipole subtraction for non-collinear-safe
  observables}}, Nucl. Phys. \textbf{B800} (2008),
  \href{http://www.slac.stanford.edu/spires/find/hep/www?eprint=0802.1405}{146--189},
   [\href{http://arXiv.org/pdf/0802.1405}{{\tt arXiv:0802.1405}} [hep-ph]]%
\relax\mciteBstWouldAddEndPuncttrue
\mciteSetBstMidEndSepPunct{\mcitedefaultmidpunct}
{\mcitedefaultendpunct}{\mcitedefaultseppunct}\relax
\EndOfBibitem
\bibitem{Kallweit:2017khh}
S.~Kallweit, J.~M. Lindert, S.~Pozzorini and M.~Sch{\"o}nherr, \emph{{NLO
  QCD+EW predictions for $2\ell2\nu$ diboson signatures at the LHC}}, JHEP
  \textbf{11} (2017),
  \href{http://www.slac.stanford.edu/spires/find/hep/www?eprint=1705.00598}{120},
   [\href{http://arXiv.org/pdf/1705.00598}{{\tt arXiv:1705.00598}} [hep-ph]]%
\relax\mciteBstWouldAddEndPuncttrue
\mciteSetBstMidEndSepPunct{\mcitedefaultmidpunct}
{\mcitedefaultendpunct}{\mcitedefaultseppunct}\relax
\EndOfBibitem
\bibitem{Amati:1980ch}
D.~Amati, A.~Bassetto, M.~Ciafaloni, G.~Marchesini and G.~Veneziano, \emph{{A
  Treatment of Hard Processes Sensitive to the Infrared Structure of QCD}},
  Nucl. Phys. B \textbf{173} (1980),
  \href{http://www.slac.stanford.edu/spires/find/hep/www?j=Nucl%20Phys%20B,173,429}{429--455},
  CERN-TH-2831%
\relax\mciteBstWouldAddEndPuncttrue
\mciteSetBstMidEndSepPunct{\mcitedefaultmidpunct}
{\mcitedefaultendpunct}{\mcitedefaultseppunct}\relax
\EndOfBibitem
\bibitem{Brodsky:1982gc}
S.~J. Brodsky, G.~P. Lepage and P.~B. Mackenzie, \emph{{On the Elimination of
  Scale Ambiguities in Perturbative Quantum Chromodynamics}}, Phys. Rev. D
  \textbf{28} (1983),
  \href{http://www.slac.stanford.edu/spires/find/hep/www?j=Phys%20Rev%20D,28,228}{228},
  SLAC-PUB-3011, FERMILAB-PUB-83-040-T%
\relax\mciteBstWouldAddEndPuncttrue
\mciteSetBstMidEndSepPunct{\mcitedefaultmidpunct}
{\mcitedefaultendpunct}{\mcitedefaultseppunct}\relax
\EndOfBibitem
\bibitem{Cacciari:2008gp}
M.~Cacciari, G.~P. Salam and G.~Soyez, \emph{{The Anti-k(t) jet clustering
  algorithm}}, JHEP \textbf{04} (2008),
  \href{http://www.slac.stanford.edu/spires/find/hep/www?eprint=0802.1189}{063},
   [\href{http://arXiv.org/pdf/0802.1189}{{\tt arXiv:0802.1189}} [hep-ph]]%
\relax\mciteBstWouldAddEndPuncttrue
\mciteSetBstMidEndSepPunct{\mcitedefaultmidpunct}
{\mcitedefaultendpunct}{\mcitedefaultseppunct}\relax
\EndOfBibitem
\bibitem{Buckley:2010ar}
A.~Buckley et~al., \emph{{Rivet user manual}}, Comput. Phys. Commun.
  \textbf{184} (2013),
  \href{http://www.slac.stanford.edu/spires/find/hep/www?eprint=1003.0694}{2803--2819},
   [\href{http://arXiv.org/pdf/1003.0694}{{\tt arXiv:1003.0694}} [hep-ph]]%
\relax\mciteBstWouldAddEndPuncttrue
\mciteSetBstMidEndSepPunct{\mcitedefaultmidpunct}
{\mcitedefaultendpunct}{\mcitedefaultseppunct}\relax
\EndOfBibitem
\bibitem{Bierlich:2019rhm}
C.~Bierlich et~al., \emph{{Robust Independent Validation of Experiment and
  Theory: Rivet version 3}}, SciPost Phys. \textbf{8} (2020),
  \href{http://www.slac.stanford.edu/spires/find/hep/www?eprint=1912.05451}{026},
   [\href{http://arXiv.org/pdf/1912.05451}{{\tt arXiv:1912.05451}} [hep-ph]]%
\relax\mciteBstWouldAddEndPuncttrue
\mciteSetBstMidEndSepPunct{\mcitedefaultmidpunct}
{\mcitedefaultendpunct}{\mcitedefaultseppunct}\relax
\EndOfBibitem
\bibitem{Basso:2015gca}
L.~Basso, S.~Dittmaier, A.~Huss and L.~Oggero, \emph{{Techniques for the
  treatment of IR divergences in decay processes at NLO and application to the
  top-quark decay}}, Eur. Phys. J. \textbf{C76} (2016), no.~2,
  \href{http://www.slac.stanford.edu/spires/find/hep/www?eprint=1507.04676}{56},
   [\href{http://arXiv.org/pdf/1507.04676}{{\tt arXiv:1507.04676}} [hep-ph]]%
\relax\mciteBstWouldAddEndPuncttrue
\mciteSetBstMidEndSepPunct{\mcitedefaultmidpunct}
{\mcitedefaultendpunct}{\mcitedefaultseppunct}\relax
\EndOfBibitem
\bibitem{Dittmaier:1999mb}
S.~Dittmaier, \emph{{A general approach to photon radiation off fermions}},
  Nucl. Phys. \textbf{B565} (2000),
  \href{http://www.slac.stanford.edu/spires/find/hep/www?eprint=hep-ph/9904440}{69--122},
   [\href{http://arXiv.org/pdf/hep-ph/9904440}{{\tt arXiv:hep-ph/9904440}}
  [hep-ph]]%
\relax\mciteBstWouldAddEndPuncttrue
\mciteSetBstMidEndSepPunct{\mcitedefaultmidpunct}
{\mcitedefaultendpunct}{\mcitedefaultseppunct}\relax
\EndOfBibitem
\end{mcitethebibliography}
\end{document}